\documentclass[11pt]{article}

\usepackage[utf8]{inputenc}
\usepackage{cite}
\usepackage{dsfont}
\usepackage{amsmath}
\usepackage{amsfonts}
\usepackage{amssymb}
\usepackage{amsthm}
\usepackage{mathtools}
\usepackage{graphicx}
\usepackage{balance}
\usepackage{color}
\usepackage{xurl}
\usepackage{algorithmic}
\usepackage{subfigure}
\usepackage[boxed]{algorithm2e}
\usepackage{multirow}
\usepackage{float}
\usepackage{lipsum}

\newcommand{\nop}[1]{}

\begin{document}
\begin{sloppy}

\title{\bf A Survey on Data Pricing:\\ from Economics to Data Science
}
\author{Jian Pei\\
E-mail: jpei@cs.sfu.ca
}


\maketitle

\begin{abstract}
Data are invaluable. How can we assess the value of data objectively, systematically and quantitatively? Pricing data, or information goods in general, has been studied and practiced in dispersed areas and principles, such as economics, marketing, electronic commerce, data management, data mining and machine learning. In this article, we present a unified, interdisciplinary and comprehensive overview of this important direction. We examine various motivations behind data pricing, understand the economics of data pricing and review the development and evolution of pricing models according to a series of fundamental principles. We discuss both digital products and data products.  
We also consider a series of challenges and directions for future work. 
\end{abstract}

\section{Introduction}

In this digital economics era, data are well recognized as an essential resource for work and life.  Many products and services are delivered purely in digital forms. Many big data applications are built on the second use or reuse of data~\cite{vandeSandt19}, that is, the same data are customized and reused by many applications for different purposes.  The extensive sharing and reusing data has profound implications to economy.  For example, digital maps are often produced for traffic and directions as the immediate usage.  However, Nagaraj~\cite{Nagaraj16} finds that mining activities were strongly benefited by open maps or maps sponsored by governments, particularly for smaller firms with less resources. Universal availability of data often helps minority parties and emerging initiatives.

In business and economic activities where data are shared, exchanged and reused, it is essential to measure the value of data properly.  While there exist many possible ways to appreciate and represent the value of data, a general approach that can be scalable for massive applications and acceptable to many parties is to set a price at which data can be sold or purchased, that is, data pricing.  The importance of pricing in business is well recognized in financial modeling~\cite{Kotler00}, as price being one of the four Ps of the marketing mix\footnote{The four Ps are product, price, place and promotion~\cite{Kotler00}.}. 

Pricing data is far from trivial. Data have many different aspects.  Consequently, the term ``price of data'' may carry different meanings and refer to different properties of data.  To illustrate the complexity, let us quickly consider the following three scenarios involving price information related to data.  

\begin{itemize}

\item \emph{Data transmission.} Imagine the scenario where a mobile service provider offers a smart phone user the price of its data package.  Here, the price is quoted for the data transmission service and is decided by several factors, such as the amount of data the user wants to transmit in a month time, the location (roaming or not, for example), and the transmission speed.  The price does not include and is independent from the content, that is, what the data are about, such as data quality, and how the data are collected, stored or processed.

\item \emph{Digital products.} Imagine that a person wants to watch a movie at home.  This is a purchase of data, since the movie is sent to the customer's home as a stream of bits. The price here typically is related to the content, but is independent from the data transmission service, that is, how the data are transmitted to the user's home.

\item \emph{Data products.}  Many logistics companies want to pay for weather information to support their business operations.  While historical data are relevant, more often than not those companies want to subscribe to weather forecasting information instead. Some companies may want weather predictions at a higher granularity while some may want detailed predictions at specific locations.  Moreover, some may want long term predictions while some others may want short term projections.  Here, prediction services are sold as data products.

\end{itemize}

The above three cases just elaborate some representative scenarios where data prices are used, and are by no means exhaustive. To appreciate data pricing, including ideas, principles and methods, we have to take an interdisciplinary approach from multiple fields, economics and data science being the two most prominent. Indeed, the studies and practice of data pricing started as early as the dawn of digital economics, and are highly diversified and rich in innovative thinking.

In this article, we try to present a comprehensive survey on data pricing, an emerging research and practice area that plays a more and more important role in the current big data and AI economics era.  Our survey is highly related to the current strong rising of data science.   To a large extent, data pricing is an overdue pillar in data science research and practice.

Data and information as goods discussed in this article are those that are distributed purely in digital form.  We focus on two categories of the most interest: pricing digital products and pricing data products, demonstrated by the last two aforementioned scenarios, respectively. In this article, \emph{digital products} refer to those intangible goods but can be consumed through electronics, such as e-books, downloadable musics, online ads, and internet coupons. Many digital products have physical correspondences in one way or another, though not absolutely necessary. \emph{Data products} refer to data sets as products and information services derived from data sets. We build the linkage between these two categories by pointing out many ideas and methods on pricing digital products can be generalized and applied to pricing data products. In some scenarios, the boundary between digital products and data products is also blurry.  Hereafter, we use the term \emph{information goods} to refer to both digital products and data products.  

\subsection{Related Surveys}

The research into data pricing happens simultaneously in multiple domains, including but not limited to economics, marketing, e-commerce, databases and data management, operational research, management science, machine learning and AI.  However, to the best of our knowledge, there exists very limited effort to provide an interdisciplinary survey of the related work.  This article presents our endeavor to produce a comprehensive picture.

There are some previous surveys related to data pricing.  For example, Liang~\textit{et~al.}~\cite{LiangYuAn18} survey the life cycle of big data, and reviews 11 data pricing models.  They also discuss data trading and protection.  Fricker and Maksimov~\cite{10.1007/978-3-319-69191-6_4} report a literature survey over 18 research articles regarding several research questions, including maturity of the pricing models.  Very recently, Zhang and Beltr\'an~\cite{ZhangBletran20} review the state-of-the-art data pricing methods.  They categorize data pricing methods according to two important data properties, granularity and privacy,  This article covers a substantially broader scope than those~\cite{LiangYuAn18, 10.1007/978-3-319-69191-6_4, ZhangBletran20}.  We connect economics, digital product pricing and data product pricing.  We also discuss a series of desirable properties in data pricing, including arbitrage-freeness, revenue maximization, fairness, truthfulness, and privacy preservation, and review the techniques achieving those properties. 

Data pricing is related to cloud pricing, since a lot of data for pricing and trading are hosted on cloud.  Wu~\textit{et~al.}~\cite{10.1145/3342103} present a comprehensive survey on cloud pricing models. They systematically categorize three fundamental pricing strategies, namely value-based pricing, cost-based pricing and market-based pricing.  Then, they further categorize nine pricing tactical objects. Specifically, value-based pricing is demand driven and consists of customer value-based pricing, experience-based pricing, and service-based pricing.  Cost-based pricing is supply driven and consists of expenditure-based pricing, resource-based pricing and utility-based pricing.  Market-based pricing is an equilibrium of supply and demand and consists of free and pay later pricing, retail-based pricing and auction and online pricing.  They cover in total 60 pricing models. While data and cloud are highly related, data pricing and cloud pricing are fundamentally different.  Data pricing is selling data, while cloud pricing is selling cloud resources (e.g., storage and computation), including physical resources, virtual resources and stateless resources.  

In addition, Sen~\textit{et~al.}~\cite{10.1145/2543581.2543582} survey the major broad-band pricing proposals, including the realizations in various consumer data plans around the world.  Murthy~\textit{et~al.}~\cite{10.1145/2345396.2345421} list different pricing models and pricing schemes used by some popular IaaS (infrastructure-as-a-service) providers. Wu~\textit{et~al.}~\cite{6392064} propose pricing as a service, which is essentially a personalized pricing service for IaaS.  Aazam and Huh~\cite{7037703} propose broker as a service, which matches cloud services among cloud service providers and users. The key idea is to predict resource demands and thus derive prices.

As data are often hosted online, one interesting question is the fair sharing of the cost among data owners, data users and brokers.  This is related to data pricing, because the costs of data hosting and processing have to be recovered from data pricing. Kantere~\textit{et~al.}\cite{10.1145/1989323.1989358} study the fair allocation of costs in query services. They develop a stochastic model, which predicts the extent of cost amortization in time and number of services based on query traffic statistics. The model can be implemented on top of a cloud DBMS. Al-Kiswany~\textit{et~al.}~\cite{10.1145/2452376.2452447} provide a cost assessment tool to evaluate the cost of a desired data sharing.  One useful feature of the tool is that a user can explore the cost space of alternative configurations using various factors, such as quality, staleness, and accuracy. The technique is based on what-if analysis.

\subsection{Structure of This Survey}

We take a multi-disciplinary approach in this survey. The rest of the article is organized as follows.

In Section~\ref{sec:economics}, we start from economics and focus on two aspects. First, we discuss cost reduction in information goods that contributes to their prices and has impact on economics.  Then, we discuss the differences between digital products and data products.

In Section~\ref{sec:principles}, we discuss the fundamental principles of data pricing. We first present versioning as a general framework for pricing information goods.  Then, we identify several desirable properties in data pricing, including truthfulness, fairness, revenue-maximization, arbitrage-freeness, privacy preservation and computational efficiency.

In Section~\ref{sec:digitalproducts}, we discuss pricing digital products. We first 
review the three major streams of revenues for digital products.  Then, we revisit the bundling and subscription planning pricing models. Last, we consider auctions, which are widely used in pricing digital products. 

In Section~\ref{sec:data-products}, we discuss pricing data products.  We first overview the structures, players, and ways to produce data products in data marketplaces.  Then, we examine several important areas in pricing data products, including arbitrage-free pricing, revenue maximization pricing, fair and truthful pricing, privacy preservation in pricing. We also discuss dynamic data pricing, online pricing, and pricing in federated and collaborative learning.  

Last, in Section~\ref{sec:discussion}, we discuss challenges and future directions.

\section{Economics of Data Pricing}\label{sec:economics}

In general, \emph{pricing} is the practice that a business sets a price at which a product or a service can be sold.  Pricing is often part of the marketing plan of a business.  To set prices, a business often considers a series of objectives, such as profitability, fitness in marketplace, market positioning, price consistency across categories and products, and meeting or preventing competition. Some major pricing strategies in literature~\cite{oro2041, 2135, Brennan13, 83820, Irvin79} include operation-oriented pricing, revenue-oriented pricing, customer-oriented pricing, value-oriented pricing, and relationship-oriented pricing. There is a rich body of studies in economics and marketing research on pricing tactics, which are far beyond the scope and capacity of this survey.

In this section, to understand the economic factors specific to data pricing, we examine the cost reduction in information goods.  Then, we inspect the differences between digital products and data as products.  

\subsection{Cost Reduction in Information Goods}\label{sec:cost-reduction}

``Technology changes. Economic laws do not.''~\cite{1998information} The production, distribution, and consumption of information goods, comparing to those of physical products in the long history of human economies, are distinguished by significant cost reductions on five aspects, namely search costs, production costs, replication costs, transportation costs, and tracking and verification costs. Essentially, digital and data economics investigates how standard economic models adjust when those costs are reduced dramatically. Goldfarb and Tucker~\cite{10.1257/jel.20171452} present a thorough discussion, whose framework is largely followed here.

\subsubsection{Search Costs}

``\emph{Search costs} are the costs of looking for information''~\cite{1998information}, which are incurred in any information collection activities.  Information goods allow more effective and efficient online search.  The consequent low search costs facilitate users' discovering digital products and data sets, as well as comparing prices of similar products and services.  For example, Brynjolfsson and Smith~\cite{doi:10.1287/mnsc.46.4.563.12061} show that online prices of books and CDs are clearly lower than offline, though the price dispersion, however, does not shrink accordingly.

Low search costs facilitate the sales of rare and long tail products~\cite{RePEc:bla:randje:v:44:y:2013:i:4:p:733-756, 10.5555/1197299}. Thus, more variety is often observed in information goods and services. The degree of variety may be heavily impacted by recommender systems.  Specific to consumption of media, one of the major categories of digital products, Gentzkow and Shapiro~\cite{10.1093/qje/qjr044} show that online media consumption is more diverse than offline.  At the same time, customers may tend to consume more that aligns more or less with their viewpoints, which is called the ``echo chamber'' effect~\cite{sunstein2001echo}.

Low search costs give strong rise to the prevalent platform businesses, which provide extensive matching services to customers and improve trade efficiency~\cite{TwoSidedBtoBPlatforms}. Interoperability, compatibility and standards are strategic tools for both building platforms and running platform businesses~\cite{10.2307/43189630}.

\subsubsection{Production Costs}

Producing digital products, such as online courses, eBooks, software, graphics and digital arts, and photography, is very different from manufacturing physical products, like bread, shoes, and jackets. Moreover, collecting and processing massive data so that parts of data can be sold and can meet customers' needs is also different from traditional production.  A wide spectrum of production costs in traditional products are substantially reduced in information goods.

First, some essential major costs in traditional production, such as materials, semi-finished products and their transportation, are dramatically reduced in producing information goods.  In many cases, the costs of obtaining, producing and transporting raw materials and physical semi-finished products can be reduced to very low or can even approach zero in making information goods.
Second, a substantial cost of a traditional physical product often belongs to the product itself and cannot be further reduced through sharing.  The unit costs of information goods can approach zero through sharing as long as there are sufficient reuses and sales volume. 
Last, smart manufacturing and customer-to-manufacturing can reduce the supply chain costs in traditional physical production~\cite{10.5555/2994178, SBRB09}.  Information goods often can reduce the costs of customization to extreme.  

The substantial reduction in production cost in materials, semi-finished products, customization and sharing gives rise to a series of innovative business models, such as economics of sharing, pay-as-you-go and query-based data consumption. This also encourages innovation and long tail products that address diverse and smaller groups of potential customers.

\subsubsection{Replication Costs}\label{sec:replication-cost}

One distinct feature of information goods versus traditional products is that information goods are non-rival.  That is, one customer consuming an information good does not reduce the amount or quality of the product available to other customers.  The zero marginal costs and the non-rival property of information goods empower innovative opportunities and bring in new challenges.

In order to structure pricing of a large variety of non-rival information goods with zero marginal costs, bundling is often used~\cite{1998information}, that is, multiple products are sold together at a single price. Since a large number of information goods can be bundled together without a substantial increase in cost, economically it may be optimal to bundle thousands of digital products together to meet diverse and independent customer preferences~\cite{10.1287/mnsc.45.12.1613, 10.5555/2882592.2882597, RePEc:eee:indorg:v:57:y:2018:i:c:p:278-307}.

Due to the zero marginal costs and the non-rivalrous property, many information goods are made publicly available, such as Wikipedia\footnote{\url{https://www.wikipedia.org}} and open source software~\cite{10.1257/000282806777211874}. People contribute to open source or publicly available digital products and data to demonstrate their professional skills to potential employers.  Companies support those products to complement their sales on other products. 

The zero marginal costs and non-rivalrous property post challenges to copyright policies and enforcement. Waldfogel~\cite{10.1257/aer.102.3.337} shows that low replication costs, though may reduce revenue, help supplies and demands, and thus boost quality. Williams~\cite{doi:10.1086/669706} shows that the protection of intellectual properties indeed has negative impact on follow-on innovation in gene sequencing.

At the same time, there are evidences showing that governments mandate ``open data'' may lead to data leakages and privacy breaches that affect citizens' offline welfare~\cite{Acquisti11gunsprivacy}.  On the negative side, the zero marginal costs or non-rivalrous nature also ease the way for spamming~\cite{10.1257/jep.26.3.87} and online crime~\cite{10.1257/jep.23.3.3}.

\subsubsection{Transportation Costs}

Thanks to the Internet, the costs of transporting information goods approach zero.  This may imply, in many scenarios, that local communities may not affect adoptions and consumptions of information goods, often known as the effect of flat world~\cite{flatworld}. Interestingly, this is not true all the time, as some studies demonstrate that tastes may still be local in music~\cite{21f58d6f112f49b88a47e4ef1cf2c48f} and content consumption~\cite{NativelanguageandInternetusage}.

While the physical transportation may approach zero, regulation may put sophisticated constraints on locations.  For example, when Wekipedia was blocked in China in October 2005, more contributors from outside China were motivated to contribute~\cite{10.1257/aer.101.4.1601}. Copyright policies may also affect the availability and consumption of information goods in different regions, such as news media~\cite{doi:10.1111/jems.12207}, and thus may be reflected by price.

\subsubsection{Tracking and Verification Costs}\label{sec:track-cost}

The capability of tracking users with relatively low costs is an important feature of information goods~\cite{1998information}. The low tracking costs give the rise to extensive personalized markets and possible price discrimination~\cite{10.1145/948005.948051, PriceDiscrimination}. Behavioral price discrimination is an immediate type, which sets prices according to customers' previous behavior. Correspondingly, if customers are well aware of the benefits of tracking information to a monopoly, they may likely choose to be privacy sensitive and hold the information~\cite{RePEc:rje:randje:v:35:y:2004:4:p:631-650}.  Another type of price discrimination is versioning~\cite{Versioning98}, which sells information at different prices to different customers using different versions. Versioning is discussed in detail in Section~\ref{sec:versioning}.

The advantage of low tracking costs also leads to the blooming businesses of personalized advertising~\cite{10.1257/jep.23.3.37}. A challenge for a company, however, is how to set prices for many advertisements that may be shown to massive customers?  The same advertisement may have different prices for different customers.  Auctions are often used to address the challenge~\cite{10.1145/2764468.2764537}, and can even be used to discover prices for information goods~\cite{NBERw12785}. At the same time, auctions may be less useful when online marketplaces become mature~\cite{RePEc:ucp:jpolec:doi:10.1086/695529}. 

The low tracking costs and the consequences, such as price discrimination, lead to serious concerns on privacy~\cite{10.1257/jel.54.2.442}. As to be discussed later in this article, whether privacy should be treated as goods and how privacy is priced are investigated~\cite{10.1145/2229012.2229054, 10.1145/3219819.3220013}.  Moreover, privacy regulation and the impact on welfare are important topics, though they are far beyond the scope of this survey.

As a byproduct of low tracking costs, the costs of verifying identity and reputation of producers and users of information goods are dramatically lower than those in traditional scenarios. The low verification costs facilitate online transactions extensively and lower the costs of trust dramatically.

\subsection{Differences between Digital Products and Data Products}\label{sec:differences}

This survey focuses on pricing two categories of information goods, digital products and data products.  While digital products and data products share a series of common ideas and methods in pricing, they are also essentially different from each other on at least four aspects.

First, the units of digital products are often well defined and fixed.  For example, individual movies and musics are often priced and sold in whole.  The consumption of a digital product is often independent from each other.  For example, it would be rare that two digital books have to be read at the same time.  In contrast, although the basic unit in a data set can be at a very small granularity, such as a record in a relational table, the units for pricing and consumption often vary from one customer to another.  For example, a customer may be interested in the sales data of female customers in a province, while another customer may be interested in the sales data on electronics during the Christmas season. Correspondingly, one individual unit of data at the lowest granularity may not be valuable as a data product.  For example, one customer purchase record, after proper anonymization, may not be useful for a retailer.  Instead, more often than not, many basic units of data are combined, aggregated and consumed together.

Second, different from digital products, data sets as data products have very strong and flexible aggregateability.  Customers often aggregate data using various dimensions. The aggregateability, on the one hand, enables many opportunities for innovations in data business, and, on the other hand, posts many technical and business challenges, such as ensuring arbitrage-freeness as to be discussed later in this article.  In many business scenarios, digital products like movies and musics are bundled.  However, bundles are not aggregates.  Customers still get digital products and consume them individually.  Bundling is to take the advantage of low replication costs of digital products to boost sales and meet customers' diverse demands~\cite{10.1287/mnsc.45.12.1613, 10.5555/2882592.2882597, RePEc:eee:indorg:v:57:y:2018:i:c:p:278-307}.

Third, the means of consuming digital products and data products are also very different. Typically digital products are consumed directly by people, such as movies watched by people and musics enjoyed by fans.  Data sets are more often than not consumed by computers.  They are, for example, analyzed, summarized or used to train machine learning models.  The outputs of models are used to automate operations or support human decision making.  

Last, digital products and data products are dramatically different in ways to be reused and resold. Digital products are easy to be consumed by others, that is, to be reused, or even to be resold to others in whole.  Data sets, to the contrary, can be reused by others in different ways, such as aggregation in different dimensions and analysis for different purposes.  Moreover, data can be easily processed and transformed so that they can be resold in a hard-to-detect manner.

The above differences between digital products and data products lead to different considerations in pricing principles and methods, which are discussed later.  Before we leave this topic, we want to point out that it is possible that the same information can be regarded as digital products in some situations and as data products in some other situations.  For example, social media like tweets and customer reviews can be regarded as digital products when a customer reads them online.  At the same time, they can be collected and processed in batch by analytic tools to detect events, discover customer profiles and feed recommender systems. In this situation, a systematic collection of social media can be priced and sold as a data product.

\subsection{Summary}

In summary, information goods, including digital products and data products, distinguish themselves from the traditional physical products in significant cost reductions, particularly in search costs, production costs, replication costs, transportation costs, and tracking and verification costs.  The significant reduction of costs has profound impact on pricing information goods, which is discussed in the later sections of this article.  There are several major differences between digital products and data products, including consumption units, aggregatebility, means of consumption, and reusing and reselling.


\section{Fundamental Principles of Data Pricing}\label{sec:principles}

In this section, we first review the idea of versioning~\cite{Versioning98, 1998information}, which is a fundamental framework of designing information goods and pricing them.  Then, we review several important properties in cost models of digital and data products. 

\subsection{Versioning}
\label{sec:versioning}

As the replication costs of information goods are very low, even approaching zero in many cases, the price of an information good tends to be very low in marketplaces, too. The potential of very low prices of information goods, on the one hand, makes information goods economically appealing, and, on the other hand, may also make information goods economically dangerous, as the competitors may easily enter the market~\cite{Versioning98, 1998information}.  This dilemma keeps many traditional pricing strategies far away from being effective for information goods.

To tackle the dilemma, the core idea is ``linking price to value'', that is, setting the price reflecting the value that a customer places on the information.  Specifically, the \emph{versioning strategy}~\cite{Versioning98} makes different versions to appeal to different types of customers. For example, for a piece of software, different versions have different subsets of features.  Different versions of a movie may provide different image resolutions and sound effects.  Essentially, versioning divides customers into subgroups so that each subgroup may regard some features highly valuable and some other features of little value.  A version corresponding to the demands can be provided.

There are many different ways to produce different versions of information goods.  For example, as information is often time sensitive, delay is often a good basis.  In stock market information services, an expensive version may deliver real time quotes while a basic version delivers the same information 20 minutes later. In addition, versions may be defined by convenience (e.g., data can be accessed only by PDF file or by downloadable spreadsheet), comprehensiveness (e.g., the length of historical data available), manipulation (e.g., whether users can store, duplicate, print the information), community (e.g., availability of posting and reading discussion boards), annoyance (e.g., the option of no advertisements), the means of customer support (e.g., by website only or by talking to experts), and many other factors.  Most versions of information goods are created by subtracting value from the most technologically advanced and complete version.

In many situations where customers may not realize the value of an information good unless they try it, even the free versions may be provided. The rationale is that the free versions can provide opportunities to potential customers to test out.  The objectives of offering free versions include building awareness, gaining follow-on sales, creating a customer network, attracting attentions, and gaining competitive advantages.

The number of versions of an information good may be decided by two major considerations. First, the characteristics of the information to be sold is important.  An information good that can be used in many different ways opens the door to many different versions. The second important factor is the value that different customers may place on it.  The larger the variance, the more versions may be needed.

The versioning strategy has been investigated in pricing data products, for example, relational data sets and query results~\cite{journals/pvldb/BalazinskaHS11, Balazinska2013}.  Relational views provide a natural and flexible technical mean to produce versions of an information source. A series of technical challenges are identified, such as arbitrage in pricing, fine-grained data pricing, pricing updates, integrated data and competing data sources, which are reviewed further in this article.

\subsection{Important Desiderata in Data Pricing}

There are many different ways to design and implement pricing models for information goods.  There are a small number of desiderata pursued by most models. How to implement those desiderata in pricing models is discussed in the later sections.

\subsubsection{Truthfulness}

To make a market efficient, the market is preferred to be truthful.  A market is truthful if every buyer is selfish and only offers the price that maximizes the buyer's true utility value.  In other words, in a truthful market, no buyer pays more than sufficient to purchase a product. Here, different buyers may have different utility values on the same product. Truthfulness can facilitate a wide spectrum of pricing mechanisms, such as many kinds of auctions~\cite{10.1145/3328526.3329589}.  Auctions of digital products are discussed in Section~\ref{sec:auctions}.

\subsubsection{Revenue Maximization}\label{sec:revenuemax}

Pricing models can optimize different objectives, such as lowest cost, highest profit, and largest sales.  The objective of maximizing revenue is often of special interest in designing pricing strategies. The rationale is that, for a business to be successful long term, a more immediate and important requirement is to win over as many customers as possible. 

For traditional physical products, it is often assumed that the marginal cost goes up after a certain number of units are manufactured, and thus the profit can be maximized if the output level is set so that the marginal revenue is equal to the marginal cost, and the revenue can be maximized if the marginal revenue becomes zero.  However, given that the replication costs of information goods are very low, revenue maximization and profit maximization for information products become quite different from those for physical products~\cite{10.1145/3328526.3329589, 10.14778/3357377.3357378}.

\subsubsection{Fairness}

Essentially, a market is fair if each seller gets the fair share of the revenue in coalition. In his seminal article~\cite{Shapley}, Shapley lays out the fundamental requirements of fairness in markets.  Suppose there are $k$ sellers cooperatively participate in a transaction that leads to a payment $v$.  There are four basic requirements for being fair.

\begin{itemize}
  \item \emph{Balance}: the sum of the payment to each seller should be equal to $v$.  That is, the payment is fully distributed to all sellers.
  \item \emph{Symmetry}: for a set of sellers $S$ and two additional sellers $s$ and $s'$ who are not in $S$, that is, $s, s' \not\in S$, if $S \cup \{s\}$ and $S \cup \{s'\}$ produce the same payment, then $s$ and $s'$ should receive the same payment.  That is, the same contribution to utility should be paid the same.
  \item \emph{Zero element}: for a set of sellers $S$ and an additional seller $s \not\in S$, if $S \cup \{s\}$ and $S$ produce the same payment, then $s$ should receive a payment of $0$.  That is, no contribution, no payment.
  \item \emph{Additivity}: If the goods can be used for two tasks $T_1$ and $T_2$ with payment $v_1$ and $v_2$, respectively, then the payment to complete both tasks $T_1 + T_2$ is $v_1 + v_2$. 
\end{itemize}

In the above well celebrated Shapley fairness, the \emph{Shapley value} is the unique allocation of payment that satisfies all the requirements.
\begin{equation}\label{eq:shapley}
\psi(s) = \frac 1 n \sum_{S \subseteq D \setminus \{s\}} \frac{\mathcal{U}(S \cup (s))- \mathcal{U}(S)}{\binom{n-1} {|S|}}
\end{equation}
where $\mathcal{U}()$ is the utility function, $D$ is the complete set of sellers, $S \subseteq D$ is a set of sellers, and $s$ is a seller.

Equivalently, Equation~\ref{eq:shapley} can also be written as
\begin{equation}\label{eq:shapley-permutation}
\psi(s)=\frac 1 {N!} \sum_{\pi \in \Pi(D)}(\mathcal{U}(P^\pi_s\cup\{s\})-\mathcal{U}(P^\pi_s))
\end{equation}
where $\pi \in \Pi(D)$ is a permutation of all sellers, and $P_i^\pi$ is the set of sellers preceding $s$ in $\pi$.

Agarwal \textit{et~al.}~\cite{10.1145/3328526.3329589} observe that, as the replication costs of information goods are very low, the marginal costs of production are close to zero, a seller can produce more units of the same information good to obtain a larger Shapley value and thus a larger portion of the payment unjustified in business. This is a challenge in designing fair marketplace for information goods.

\subsubsection{Arbitrage-free Pricing}\label{sec:arbitrage-definition}

Arbitrage is the activities that take advantage of price differences between two or more markets or channels.  For example, consider a scenario where a user wants to purchase the access to an article, whose listed price is \$35.  Suppose that the journal publishing the article has a monthly subscription rate of \$25. Then, the user can conduct arbitrage to subscribe to the journal for only one month and obtain the article at a price cheaper than the listed price.

Arbitrage is often undesirable in pricing models. At least it should be able to check whether a pricing model is arbitrage-free.  However, arbitrage can sneak in pricing models that are not thoroughly designed.  For example, suppose a data service provider sells query results with prices based on variance~\cite{10.1145/2691190.2691191}, a variance of 10 for \$5 each query result and a variance of 1 for \$100 each query result. Each answer is perturbed independently.  A customer who wants to obtain an answer of variance of 1 can purchase the query 10 times and compute their average.  Due to the independent noise in perturbation, the aggregated average has variance 1, and thus the customer saves \$50 by arbitrage.

\subsubsection{Privacy-preservation}

Privacy is becoming a more and more serious concern about information goods. In general, privacy is the ability of an individual or a group to keep themselves or the information about themselves hidden from being identified or approached by other people. Privacy is highly related to information and information exchange, which are what information goods about.

As explained in Section~\ref{sec:track-cost}, due to the low tracking costs of information goods, it is easier to collect data about user privacy~\cite{10.1257/jel.54.2.442}. Whether privacy should be treated as goods and how privacy is priced are investigated~\cite{10.1145/2229012.2229054, 10.1145/3219819.3220013}.

It is highly desirable to preserve privacy in marketplaces of information goods. In general, transactions in a marketplace may disclose privacy of various parties in many different ways.  

First, privacy of buyers is highly vulnerable.  Their identities, the location and time of purchases, specific products purchased, the purchase prices and total amount may reflect their privacy. It has been reported from time to time that e-commerce providers leak customer information by mistakes, such as an accident reported recently\footnote{\url{https://www.telegraph.co.uk/technology/2020/03/10/leak-millions-amazon-ebay-transactions-exposes-customer-addresses/}}.  

Second, privacy of information good providers may also be disclosed.  For example, medical treatment information in hospitals is highly valuable for many business companies, such as pharmacy and medical equipment companies. Imagine that hospitals can collect and anonymize medical treatment data properly and provide the corresponding data products in marketplaces so that individual patients cannot be re-identified. Buyers, however, may be able to infer from the data the successful rates of a specific treatment in a hospital, which may be regarded as the privacy of the hospital. 

Last, transactions in marketplaces may also disclose privacy of a third party involved.  For example, an AI technology company may provide machine learning model building services to data product buyers.  However, machine learning models may be stolen~\cite{10.5555/3241094.3241142}, which are regarded privacy of the AI technology company.

To protect privacy in marketplaces of information goods, various directions are being explored, such as hiding the information about what, when and how much a buyer purchases~\cite{eurocrypt-2001-2009}, building decentralized and trustworthy privacy preservation data marketplace~\cite{10.14778/3229863.3236266, DBLP:journals/corr/abs-1802-04780}, investigating the tradeoff between payments and accuracy when privacy presents~\cite{10.1145/2554797.2554835}, and aggregating non-verifiable information from a privacy-sensitive population~\cite{10.1145/2600057.2602902}. There are many studies on preserving privacy in information goods. We refer interested readers to consult the rich body of surveys~\cite{10.1145/1749603.1749605, Aggarwal2008, Bertino2008, 10.1007/978-3-540-79228-4_1, 10.1145/1540276.1540279, 7954609, Wu2010, Jiang2020DifferentialPA} and others.  We do not discuss further details about general privacy preservation techniques in this article, since privacy preservation techniques are far beyond the scope and capacity of this survey.

\subsubsection{Computational Efficiency}

As many information goods may be sold to a huge number of potential buyers, a pricing model has to match goods/sellers and buyers with an appropriate price.  Computing prices efficiently with respect to a large number of goods and a large number of buyers presents technical challenges~\cite{journals/pvldb/BalazinskaHS11}.  

For example, one reasonable expectation is that a marketplace is polynomial, that is, the complexity of computing prices has to be polynomial with respect to the number of sellers, and cannot grow with respect to the number of goods/buyers when prices are updated~\cite{10.1145/3328526.3329589}. When auctions are used in determining prices, auction efficiency~\cite{10.5555/365411.365768} is required to be fast, which is the time needed to process bids.

\subsection{Summary}

Versioning is a common mechanism in designing and pricing information goods, so that prices of different versions can be linked to values placed by various customer groups.  There are a series of important requirements on pricing information goods, including truthfulness, revenue maximization, fairness, arbitrage-free pricing, privacy preservation, and computational efficiency.  Those requirements post technical challenges to pricing models.

\section{Pricing Digital Products}\label{sec:digitalproducts}

Although the focus of this article is about pricing data products, we provide a brief review on pricing digital products here, since some general ideas in pricing digital products can be borrowed and extended to data products.  In some cases, the boundary between digital products and data products is even blurry.  

We first discuss the three major streams of revenues for digital products.  Then, we look at two major types of pricing models.  The first is bundling and subscription, and the second is auctions. These pricing models are popularly adopted by digital product marketplaces.

\subsection{Streams of Revenues}

As discussed in Section~\ref{sec:revenuemax}, revenue maximization often serves as the basic objective in pricing mechanisms, including pricing digital products.  Therefore, the understanding of pricing digital products can naturally start with an analysis of possible ways where revenues of digital products may come from.

Lambrecht \textit{et~al.}~\cite{Lambrecht14} summarize that there are three streams of revenues for digital products that are delivered online. 

\begin{itemize}

\item \emph{Money}. A provider can sell to customers content or, more broadly, services, such as movies and e-books. 

\item \emph{Information/privacy}. Instead of charging customers directly, a provider can collect customer information by tracking (e.g., using cookies) and sell the information about customers to generate revenues.  

\item \emph{Time/attention}. A provider can sell space in their digital products to advertisers to produce revenue.  

\end{itemize}

Often, a firm has to design a revenue model for its digital products that combine more than one revenue stream. The three streams are not independent.  Instead, they compete with each other, and thus a good tradeoff has to be settled~\cite{GALLAUGHER2001473}. On the one hand, in some situations, revenues from money stream may be increased at the cost of those from time/attention stream.  For example, customers may pay for the content and avoid ads~\cite{PRASAD200313, doi:10.1509/JMKG.72.3.014}, or convert from free versions to premium versions with fitting functions~\cite{WBH14}. On the other hand, customers may be highly price sensitive in some digital products, and thus growth in time/attention stream may be easier.  For example, an online news site experiences a dramatic loss of customer visits after introducing a paywall~\cite{RePEc:eee:iepoli:v:25:y:2013:i:2:p:61-69}. Free samples may stimulate long-term sales~\cite{RePEc:zbw:fubsbe:200428}. A possible tradeoff between money and time/attention has to be carefully designed.  

Typical approaches in revenue models of content and services~\cite{RePEc:inm:ormksc:v:34:y:2015:i:3:p:430-451} include rigid pricing (e.g., each movie is priced at a fixed price), designing pricing tiers (e.g., basic versus premium versions), setting up duration of subscription plans (e.g., 6 months of promotion period with very low subscription price) and designing freemium models. One important and unique feature in digital product consumption is micropayments, which means a customer can pay a very small amount that is typically impractical in traditional transactions using standard credit cards due to network service fees. Micropayments and subscriptions have different effects on consumer behavior~\cite{NBERw19419}.

As a concrete example of revenue models, consider pricing software products~\cite{Lehmann09}. The major parameters of pricing models include formation of price, structure of payment flow, assessment base, price discrimination, price building and dynamic strategies. The formation of price considers price determination, that is, cost-based, value-based or competition-oriented, as well as degree of interaction, unilateral versus interactive. In terms of payment flow, it may be by single payment, recurring payments or combination.  The assessment base of pricing may be usage-dependent (e.g., by transaction or time) or usage-independent (e.g., server types and GPU).

As the tracking costs of digital products are low, a firm can collect customer personal data and sell such data for revenue, that is, generating revenues from information/privacy stream. Typically, personal data may include customers' identities, behavior patterns, preferences and needs.  There are various ways to sell customer data, which are also discussed in Section~\ref{sec:data-products} when data products and their marketplaces are discussed.  For example~\cite{42da603e8e64425b9bba98ef13b75b97, BestBrian04}, a website can provide direct marketing companies user activity information.  Moreover, websites can also collaborate with data management platforms (DMP, for advertising)~\cite{10.14778/2536222.2536238} and produce revenues by facilitating businesses to identify audience segments. For example, the information about how customers are connected in social networks can be used to design customized discounts in marketing campaigns~\cite{10.1145/2882903.2882961}.  Bergemann and Bonatti~\cite{10.1257/mic.20140155} develop a model of pricing customer-level information such that the data about each customer are sold individually and individual queries to the database are priced linearly. As new technologies of customer tracking become available, more pricing models may emerge. 

We want to point out that selling customer data, though serves the purpose of selling digital products, crosses the boundary between selling digital products and data products.  We review some studies on setting prices for customer data and privacy information in the next section.

To produce revenues from time/attention stream, many digital product producers and service providers embed advertisements in their products in one way or the other, and obtain remarkable or even dominant advertising income.  However, as John Wanamaker (1838-1922) wisely said, ``Half the money I spend on advertising is wasted; the trouble is I don't know which half.'' It is well recognized that it is hard to accurately measure advertising effects~\cite{LewisRao13, doi:10.1287/mksc.2018.1135}. Advertisers customize ads for online display~\cite{RePEc:eee:joinma:v:30:y:2015:i:c:p:46-55, 6729558}.

One feasible way to improve advertising effectiveness is to combine user information and advertising opportunities. Retargeted advertising~\cite{doi:10.1509/jmr.11.0503} is such an approach, which combines customer online and offline behavior data and makes firms focus on customers showing prior interest in the related products. For example, Athey \textit{et~al.}~\cite{Athey2018TheIO} consider customers with multiple homes and investigate the advertising strategies and effectiveness.

In summary, digital product and service suppliers produce revenues through three major streams, money, information/privacy and time/attention. Orthogonally, a firm can bundle its digital products and also design subscription plans that provide products and services in a specific period for a price, which is discussed next.

\subsection{Bundling and Subscription Planning}

Product bundling organizes products or services into bundles, such that a bundle of products or services are for sale as one combined product or service package.  Product bundling is a common marketing practice, particularly in the traditional industry like telecommunication services, financial services, healthcare, and consumer electronics.

As discussed in Section~\ref{sec:replication-cost}, the low replication costs of information goods allow prevalent adoption of bundling in pricing digital products~\cite{1998information}.  Designing product bundles essentially is a combinatorial optimization problem. The basic and static setting is that a customer wants to buy either one or multiple products at a time, which is investigated well before digital products are available~\cite{10.2307/1886045}. A series of studies~\cite{MENICUCCI201533, doi:10.3982/ECTA10269, RePEc:eee:jetheo:v:148:y:2013:i:2:p:448-472} develop pricing strategies with two products under different types of bundling.  They share the basic assumption that demand for a bundle is elastic comparing to demand for individual products.  For example, Armstrong~\cite{RePEc:eee:jetheo:v:148:y:2013:i:2:p:448-472} studies the scenarios where products may be substituted or provided by separate sellers.

Bundling multiple products is analyzed, often under the independent value distribution framework~\cite{10.1287/moor.6.1.58}. Consider the situation where there are $n$ heterogeneous products for one buyer, and the objective is to maximize expected revenue.  Assume that the value distributions on products are independent.  That is, for each product $x_i$, the price that a buyer would like to pay for is an arbitrary distribution $D_i$ in range $[a_i, b_i]$, where $0 \leq a_i \leq b_i < \infty$, and those distributions $D_1, \ldots, D_i$ are independent from each other. Further assume that the buyer is additive, that is, the buyer's value for a set of products is the sum of the buyer's values of those individual products in the set. Babaioff~\textit{et~al.}~\cite{doi:10.3982/ECTA12618} show that either selling each item separately or selling all items together as a grand bundle produces at least a constant fraction of the optimal revenue. This interesting and important result allows  a simple yet effective bundling strategy: either pricing each product individually or pricing the grand bundle in the expected price. In practice, many platforms, such as Hulu and Amazon Prime Video, offer grand bundle subscription for their products.  

More recently, Haghpanah and Hartline~\cite{Haghpanah2020, 10.1145/2764468.2764498} show that grand bundle is optimal if more price-sensitive buyers consider the products more complementary.
When multiple buyers are considered, whose preferences are unknown, Balcan \textit{et~al.}~\cite{10.1145/1386790.1386802} give a simple pricing model that achieves a surprisingly strong guarantee: in the case of unlimited supplies, a random single price achieves expected revenue within a logarithmic factor for customers with general valuation functions.  This result allows great convenience in practice, that is, setting a uniform price for all products. It is easier to price a bundle of a larger number of products, since the law of large numbers allows to predict customers' valuations more accurately for a larger bundle of products~\cite{doi:10.1111/poms.12958}.

Orthogonal to bundling, subscription is to price the interactions between customers and a platform over a period of time. Subscribing customers are in general heterogeneous in both usage rate and value of products. On the one hand, customers with higher usage rates may prefer subscribing to larger subscription sets. On the other hand, in order to maximize revenue, the platform wants customers with lower usage rates to subscribe, and customers with higher usage rates to rent.  Moreover, different users may have different values for a product.  Many platforms offer subscription and renting at the same time. For a platform, the \emph{subscription model} is to select a subscription fee and the period for each set of products and also set the rental price for each product~\cite{Alaei19}.

Alaei \textit{et~al.}~\cite{Alaei19} follow the model of grand bundle and consider grand subscription, a single rental price for the set that includes all products.  They establish the sufficient and necessary condition for the optimality of grand subscription.  They also show that subscription fees can be set proportional to the cardinality of a set of products and can achieve $\frac 1 {4 \log 2m + \log n}$ of the optimal revenue for $n$ types of customers and $m$ types of products.  This approximation is tight in the sense that it cannot be improved more than $\Omega(\frac 1 {\log n})$ in polynomial time.

After all, modeling bundling and subscriptions is computationally challenging due to the combinatorial nature. Dynamic pricing bundles and subscriptions, such as promotions and coupons, have rarely been touched yet.

\subsection{Auctions}\label{sec:auctions}

Auctions have a long history back to the Babylonian and Roman empires~\cite{Shubik83}. There are many excellent surveys on auctions (e.g.,~\cite{doi:10.1111/1467-6419.00083, 10.2307/2630247, 10.1257/0022051041409075, RePEc:aea:jeclit:v:25:y:1987:i:2:p:699-738}).  A comprehensive review on auctions is far beyond the scope and capacity of this article.  In this article, we instead only focus on the important role of auctions as a pricing mechanism for digital products.

\subsubsection{Basics about Auctions}

There are four basic types of auctions widely used. 

\begin{itemize}

\item In the \emph{ascending-bid auction} (also known as English auction), the price is raised successively until only one bidder remains, who wins the object at the final price. 

\item The \emph{descending auction} (also known as the Dutch auction) works the other way by starting at a very high price and lowering the price continuously, until the first bidder calls out and accepts the current price.  

\item In the \emph{first-price sealed-bid auction}, every bidder submits a bid without knowing the others' bids.  The one making the highest bid wins and pays at the named price.  

\item The \emph{second-price sealed-bid auction} (also known as the Vickrey auction~\cite{10.2307/2977633}) works in the same way as the first-price sealed-bid auction does, except that the winner pays only the second highest bid.

\end{itemize}

There are two basic models of the value information in auctions.  The \emph{private-value model} assumes that every bidder has an independent value on the object for sale.  The value is also private to the bidder only.  The \emph{pure common-value model} assumes that the actual value of the object for sale is the same for all bidders, but bidders have different private information about that actual value.  Every bidder adjusts her/his estimate of the actual value by learning other bidders' signals.  There are also models considering both values private to individual bidders and common to all bidders.

One fundamental principle in auction theory is the \emph{revenue equivalence theorem}~\cite{10.1287/moor.6.1.58, RePEc:aea:aecrev:v:71:y:1981:i:3:p:381-92, 10.2307/2977633, Vickrey62}, which essentially states that, for a set of risk-neutral bidders with independent private valuation of an object drawn from a common cumulative distribution that is strictly increasing and atomless on $[v_{min}, v_{max}]$, any auction mechanism yields the same expected revenue and thus any bidder with valuation $v$ makes the same expected payment if (1) the object is allocated to the bidder with the highest valuation; and (2) any bidder with valuation $v_{min}$ has an expected utility of $0$. Based on the revenue equivalence theorem, the four basic types of auctions lead to the same payment by the winner and the same revenue.

While most studies in auction theory make some simple assumptions about independence of customer valuations, empirical studies~\cite{10.1145/1367497.1367621} demonstrate that, in practice, the wrong assumption of valuation independence causes inefficient auctions in e-commerce.

\subsubsection{Sponsored Search Auctions}

Online ad and sponsored search auctions~\cite{lahaie_pennock_saberi_vohra_2007, 10.1257/aer.99.2.430, 10.1145/2668108} are one important application of auctions in pricing digital products.  Sponsored search~\cite{JansenMullen2008} is the business model where content providers pay search engines for traffic to their websites. In sponsored search, advertisers and, more generally, content providers bid for keywords in search engines, and search engines decide which ad to display in which position to answer a query from a user. \url{GoTo.com} created the first sponsored search auction~\cite{JansenMullen2008}.

Different pricing models can be used in sponsored search auctions, such as pay-per mille\footnote{That is, the cost of 1,000 advertisement impressions.}/pay-per impression (PPM), pay-per-click (PPC), and pay-per-action (PPA). In the early days of sponsored search, a generalized first price auction is used. Each advertiser bids on multiple keywords, and can set a bidding price for each keyword. When a user query is answered, which is a keyword, the top $k$ bids on the keyword in price are displayed.  If an ad is clicked by the user, the corresponding advertiser pays the bidding price. The first price auction mechanism is unstable, costs advertisers time and reduces search engine profits~\cite{10.1016/j.dss.2006.08.008}.  Later, Google generalizes the second price auction mechanism~\cite{10.1257/aer.97.1.242}, and enhances the ranking of bids by additional information, such as the ad's click-through-rate (CTR), keyword relevance, and ad's landing-page/site quality.

There are many in depth analyses about sponsored search auction mechanisms (e.g.,~\cite{10.1145/2668108}). For example, some studies analyze auction mechanisms based on assumptions about rationality, budget constraints and CTR distributions.  Some other studies look at practical sponsored search systems and discuss auction mechanisms when the standard assumptions do not hold.  Another group of studies, such as~\cite{CHE201720, 10.1007/978-3-540-77105-0_58, RePEc:nea:journl:y:2015:i:28:p:56-73, 10.1145/1517472.1517473}, conduct empirical studies to understand bidding behavior and statics. Last and latest, deep learning approaches are used to develop auction strategies in sponsored search~\cite{10.1145/3292500.3330870, 10.1145/3219819.3219918}.

\subsubsection{Auctions on Digital Products with Unlimited Supplies}

One unique feature of digital products is that the replication costs are very low and thus may have almost unlimited supply.  Products of unlimited supplies lead to new challenges and opportunities to auction mechanism design. For example, the second price auction can be straightforwardly generalized for $k$ identical products -- the top $k$ highest bidders win and each pays the $(k+1)$-th bidding price.  However, when there are unlimited identical products, the $(k+1)$-th bidding price approaches $0$. The lack of competition due to obsessive supplies prevents bidders from offering any high prices. In other words, the challenge is how to ensure the bids are truthful, that is, reflecting the bidders' true valuation of the digital products.

Denote by $B$ the set of bidders, and by $b_1, b_2, \ldots$ the bidding prices in descending order, that is, $b_i \geq b_{i+1} \geq 0$ for any $i > 0$.  Suppose the generalized second price auction mechanism is used.  That is, if $k$ bids are taken, those winning bidders each pays the cost $b_{k+1}$.  The auction objective is to maximize $k \cdot b_{k+1}$.  An auction is \emph{competitive} if it yields revenue within a constant factor of the optimal fixed pricing.  It is tricky that, when there is unlimited supply, the Vickrey auction is not competitive if the seller chooses the number of products to sell before knowing the bids, and is not truthful if the seller chooses after knowing the bids~\cite{10.5555/365411.365768}.

Goldberg \textit{et~al.}~\cite{10.5555/365411.365768} propose the first competitive auction for digital goods with unlimited supplies. The major idea is the smart framework of \emph{random sampling auction}. An auction is \emph{bid-independent} if bidder $i$'s bid value should only determine whether the bidder wins the auction, but not the price. We select a sample $B'$ of $B$ at random, independent from the bid values.  We use the bids in $B'$ to compute the optimal bid threshold $f_{B'}$ that maximizes the revenue in $B'$, and every bidder in $B-B'$ whose bid value is over $f_{B'}$ wins.  Symmetrically, we use the bids in $B-B'$ to compute the optimal bid threshold $f_{B-B'}$ that maximizes the revenue in $B-B'$, and every bidder in $B'$ whose bid value is higher than $f_{B-B'}$ wins.  In general, $f_{B'} = f_{B-B'}$ does not necessarily hold.  Random sampling auctions are competitive, no matter the single-price version or the multi-price version. Indeed, random sampling auctions are $15$-competitive in the worst case~\cite{10.1007/11600930_89} and $4$-competitive for a large class of instances where there are at least 6 bids that are as good as the optimal sale price~\cite{10.1145/1566374.1566402}.  There are a series of improvements on random sampling auctions.  For example, Hartline and McGrew~\cite{10.1145/1064009.1064028} further improve the competitiveness.

Goldberg and Hartline~\cite{10.5555/647911.740645} extend the scope from single digital product with unlimited supply to multiple products with unlimited supplies. Given a set of bids, they  show that the bidder-optimal product assignment given the bids and the optimal sale prices can be determined by solving the integer programming problem as follows.
\begin{equation}\label{eq:multi-item}\small
  \begin{array}{rcl}
    \max & \sum_j \sum_i x_{ij}r_j & \\
    \text{subject to} & & \\
    & r_m=0 & \\
    & \sum_j x_{ij} \leq 1 & 1 \leq i \leq n \\
    & x_{ij} \geq 0 & 1 \leq i \leq n, 1 \leq j \leq m \\
    & p_i + r_j \geq a_{ij} & 1 \leq i \leq n, 1 \leq j \leq m \\
    & \sum_i p_i = \sum_j \sum_i x_{ij}(a_{ij}-r_j) & 
  \end{array}
\end{equation}
where $x_{ij}$ is the assignment of product $j$ to bidder $i$, $r_j$ is the optimal price for product $j$, $p_i$ is the profit of bidder $i$, and $a_{ij}$ is bid from bidder $i$ on product $j$. 

Then, we can solve the optimal pricing problem in the following random sampling auction.  Let $B$ be the set of bidders.  First, we obtain a sample $B'$ of bidders. Second, we compute the optimal sale prices for $B'$.  Last, we run the fixed-price auction on $B-B'$ using the sale prices computed in Equation~\ref{eq:multi-item}.  All bidders in $B'$ lose the auction. The random sampling auction is shown truthful and competitive~\cite{10.5555/647911.740645}.

Most of the proposed auctions for digital goods with unlimited supply are randomized auctions. Goldberg \textit{et~al.}~\cite{10.5555/365411.365768} show that no deterministic auction can be competitive. Aggarwal \textit{et~al.}~\cite{10.1145/1060590.1060682} later point out that the result does not hold for asymmetric auctions~\cite{10.1111/1467-937X.00137}.  In a symmetric \emph{ex ante} auction, buyers' preference parameters are drawn from a symmetric probability distribution, and thus there exists a symmetric equilibrium if an equilibrium exists at all.  In an asymmetric auction, each buyer has the same information about the product but a different opportunity cost of obtaining the product, that is, bidders' valuations are drawn from different distributions. Aggarwal \textit{et~al.}~\cite{10.1145/1060590.1060682} give an asymmetric deterministic auction that can approximate the revenue of any optimal single-price sale in the worst case. Indeed, they develop a general derandomization technique to transform any randomized auction into an asymmetric deterministic auction with approximately the same revenue. The general idea follows the deterministic maximum flow solution to the well-known hat problem~\cite{10.5555/927911}.

\subsubsection{Envy-free Auctions}

One drawback in random sampling auctions is that some bidders may lose even they make bids higher than some winning bidders do, since the bidders in $B'$ and $B-B'$ use different thresholds (i.e., $f_{B-B'}$ and $f_{B'}$, respectively) in the one product version and all bidders in $B'$ lose in the multiple product version.

Goldberg and Hartline~\cite{10.1145/779928.779932} establish a fundamental result: an auction cannot be truthful, competitive and envy-free at the same time. They also explore possible tradeoffs between truthfulness and envy-freeness based on the consensus revenue estimate (CORE) technique~\cite{10.5555/644108.644145}. Specifically, using a similar idea in combinatorial auctions with single parameter agents~\cite{10.5555/644108.644144}, we can relax the truthfulness requirement by requiring being truthful with probability $(1-\epsilon)$, and always guarantee envy-free. The auction is highly truthful when $\epsilon$ approaches $0$ and the number of winners in the auction approaches infinity. The other type of auctions relaxes the envy-free requirement to being envy-free with probability $(1- \epsilon)$, and guarantees truthfulness.  Both auctions are competitive and the probability is over random coin tosses made by the randomized auction mechanism and not the input.

\subsubsection{Online Auctions}

In addition to potentially unlimited supply, another important feature of digital goods is that a digital good may be sold repetitively, such as a movie and a song.  Therefore, auctions on digital goods may run continuously instead of only one round.  Moreover, customers may want to have prompt answers to their bids.

Online auctions~\cite{10.1145/352871.352897} are designed to address the setting where different customers bid at different times.  The auction mechanism has to make decision about each bid as it arrives.  An (online) auction is \emph{incentive compatible} if the bidders are rationally motivated to reveal their true valuations of the object. Lavi and Nisan~\cite{10.1145/352871.352897} show that an online auction is incentive compatible if and only if it is based on supply curves under the assumption of limited supply, that is, before it receives the $i$-th bid $b_i(q)$, it fixes the supply curve $p_i(q)$ based on the previous bids, and (1) the quantity $q_i$ sold to customer $i$ is the quantity $q$ that maximizes the sum $\sum_{j=1}^q(b_i(j)-p_i(j))$; and (2) the price paid by $i$ is $\sum_{j=1}^{q} p_i(j)$. 

To tackle the challenges when there is unlimited supply, Bar-Yossef \textit{et~al.}~\cite{10.5555/545381.545506} point out that supply curves are not available anymore.  Instead, they propose an extremely simple incentive-compatible randomized online auction.  Each bidder $i$ picks a random number $t \in \{0, \ldots, \lfloor \log h \rfloor\}$ and sets the price threshold to $s_i = 2^t$, where $h$ is the ratio of the highest valuation against the lowest valuation among all bidders.  This auction is $O(\log h)$-competitive. 

The auction mechanism can be further improved to achieve even better incentive-compatibility. Specifically, we can divide a sequence of bids $b_1, b_2, \ldots$ into $l=(\lfloor \log h \rfloor+1)$ buckets, such that bucket $B_j$ contains the bids with indexes in range $[2^j, 2^{j+1})$. The weight of bucket $B_j$ is the sum of bids within $B_j$, that is, $w_j = \sum_{i \in B_j} i$. A new bidder can choose one of the buckets at random with the probability proportional to the bucket weight, and pays the price of the lowest bid of the bucket. The price $s_i$ that bidder $i$ pays follows the probability distribution
$Pr[s_i = 2^j] = \big(\frac{w_j}{\sum_{r=0}^{l-1}w_r} \big)^d$, where $d$ is a parameter. The auction is shown $O(3^d(\log h)^{\frac 1 {d+1}})$-competitive. By setting $d = \sqrt{\log \log h}$, the auction is $O(\exp(\sqrt{\log \log h}))$-competitive.

\subsection{Summary}

As revenue maximization plays a fundamental role in pricing digital products, we review the three major streams of revenues for digital products, namely money, information/privacy, and time/attention.  Then, we revisit bundling and subscription planning for digital products, which echoes the opportunities and challenges due to low replication costs of information goods. Auctions are widely used in pricing digital products.  We review some basic types of auctions and their applications in digital products, including sponsored search auctions, auctions with unlimited supplies, envy-free auctions and online auctions.  Some ideas employed by pricing digital products are also used in pricing data products, as to be discussed in the next section.

\section{Pricing Data Products}\label{sec:data-products}

In this section, we discuss pricing in marketplaces of data.  We first obtain an overall understanding about data markets and the major players in such markets. Then, we look into several most studied technical problems in data product pricing, including arbitrage-free pricing, revenue maximization pricing, fair and truthful pricing and privacy preservation in data marketplaces.  Last, we discuss pricing in novel application scenarios, including dynamic data pricing, online pricing and federated learning pricing.

\subsection{Data Markets and Pricing, What Are They?}

Marketplaces for data have been actively developed for over a decade.  An early survey~\cite{10.1145/2481528.2481532} identifies different categories and dimensions of data marketplaces and data vendors in 2012.  There are many studies on various issues about data markets and pricing strategies.  Before we discuss any specifics in detail, it is important to obtain an overall understanding about data markets, such as what are sold and for what purposes, who are the sellers, who are the buyers, and what are the basic pricing models.

Pantelis and Aija~\cite{6691691} present a brief economic analysis of data taxonomy as a market mechanism.  Data and databases are legally protected by either copyright or database right.  Copyright protects expression and significant creative effort that creates and organizes data.  Database right protects a whole database.  One challenge is that both copyright and database right are hard to enforce due to the non-rivalrous nature of data.  

In general, data may be owned by governments, private parties or individuals.  Consequently, data can be categorized into three types: open, public, and private data~\cite{6691691}.  Open data are common pool resources~\cite{ostrom_2015}, such as the data made available by the open data initiatives. Public data, such as the data collected by the government in the United States, are valuable resources subject to the ``tragedy of the commons''~\cite{Hardin1243}.  Public data are often produced by individuals or organizations for research and used by governments and local authorities, but may also be employed by commercial parties to enhance their proprietary resources or services. Private data are generated by private applications or services.

To understand what are sold in data markets and for what purposes, Muschalle~\textit{et~al.}~\cite{10.1007/978-3-642-39872-8_10} consider the common queries and demands on data markets, as well as the pricing strategies. They observe two major types of queries. The first type of queries is to estimate the value of a ``thing'' or compare the values of ``things'', where examples of the ``things'' are like webpages for advertisements, starlets, politicians and products.  The second type is to show all about a ``thing''.  Those queries are raised by seven categories of beneficiaries, namely analysts, application vendors, data processing algorithm developers, data providers, consultants, licensing and certification entities, and data market owners.  Muschalle~\textit{et~al.}~\cite{10.1007/978-3-642-39872-8_10} also identify three types of market structures. First, in a monopoly, a supplier is powerful enough to set prices to maximize profits.  Second, an oligopoly is dominated by a small number of strong competitors.  Last, in strong competition markets, prices may align with marginal costs. 

A series of pricing strategies and models may be considered in data markets~\cite{10.1007/978-3-642-39872-8_10}.  First, free data may be obtained from public authorities, may help to attract customers and suppliers of commercial data, and may be integrated into private and not-free data products.  Second, prices can be based on usages, such as charging customers per hour of data usage.  Third, package pricing allows a customer to obtain a certain amount of data or API calls for a fixed fee. A few studies~\cite{10.1145/1989323.1989358, 2235e30da74a4578a570fd9500051906} try to optimize package pricing models. Fourth, in the flat fee tariff model, a data product or service is offered at a flat rate, regardless of usage. It is simple, easy to use. The drawback is the lack of flexibility, particularly for buyers.  Fifth, combining package pricing and flat fee tariff results in two-part tariff, that is, a fixed basic fee plus additional fee per unit consumed. This model is popular in data services.  Specifically, Wu and Banker~\cite{ef53c0617eb640ed865bb5bc3da3b24c} show that, under zero marginal costs and monitoring costs, flat fee and two-part tariff pricing are on par, and two-part tariff is the most profitable strategy.  Last, in the freemium model, users can use basic products or services for free and pay for premium functions or services.

Recently, machine learning, particularly deep learning~\cite{Goodfellow-et-al-2016}, becomes disruptive in many applications, such as computer vision~\cite{Voulodimos2018, LITJENS201760} and natural language processing~\cite{8416973}. In most situations, powerful deep models heavily rely on large amounts of training data~\cite{Najafabadi2015}. Monetization of data and machine learning models built on data through markets gains stronger and stronger interests from industry.  Specific to data as an economic good and data pricing as a monetization mechanism in this context, a series of studies focus on data utility for model building and the associated pricing, particularly considering privacy. 

Some data owners may have detailed knowledge of specific machine learning tasks and thus dedicate corresponding effort to collect high quality data for building better models. Babaioff~\textit{et~al.}~\cite{10.1145/2229012.2229024} study the design of optimal mechanisms for a monopoly data provider to sell her/his data.  Specifically, they show that it is feasible to achieve optimal revenue by a simple one-round protocol, that is, a protocol where a buyer and a seller each sends a single message, and there is a single money transfer.  The optimal mechanism can be computed in polynomial time. For a buyer who may abort the interaction with a seller prematurely, multiple rounds of partial information disclosure interleaved by payments may be needed to ensure optimal revenue. Cummings~\textit{et~al.}~\cite{10.1145/2688073.2688106} study the optimal design for data buyers to purchase data estimators with different variances and combine the estimators to meet a required quality guarantee on variance with the lowest total cost.

The role of privacy in data collection and machine learning model building is investigated. For example, Ghosh and Roth~\cite{10.1145/1993574.1993605} develop auctions that are truthful and approximately optimal for data buyers to obtain accurate estimates on data from owners who are compensated for privacy loss.  They show that the classic Vickrey auction~\cite{10.2307/2977633} can minimize the buyer's total payment and meet the accuracy requirement.  They also develop a mechanism that can maximize the accuracy given a budget.  

In general, modeling data owners' costs of privacy loss is very difficult, since the costs may be correlated with private data arbitrarily.  It is impossible to design a direct revelation mechanism that can provide a non-trivial guarantee on accuracy and, at the same time, is rational for individual data owners. To tackle the issue, Ligett and Roth~\cite{LigettRoth12} design a take-it-or-leave-it mechanism, which randomly approaches individuals from a population and makes offers.  This mechanism can be used for some data collection scenarios, such as surveys.

Versioning is an important strategy in data pricing. A data seller can customize data into different versions according to buyers' needs. Bergemann~\textit{et~al.}~\cite{10.1257/aer.20161079} develop the optimal menu of information products that a monopoly data supplier can offer to a data buyer, so that one product can fit the buyer's willingness to buy the information at the offered price, and the revenue is maximized.  One important finding is that information products indeed allow larger scopes of price discrimination.  There are at least two dimensions that sellers can explore to derive various subsets of a data set, namely data quality and data position.

When data are used to build machine learning models, it is important to assess the value of each data record within a data set.  There exist various methods for assessment, such as leave-one-out~\cite{10.2307/1271434}, leverage or influence score~\cite{cook1982residuals}. 

Ghorbani and Zou~\cite{pmlr-v97-ghorbani19c} propose to apply the Shapley fairness on the data used to train a machine learning model, and thus define data Shapley for a record $i$ in a training data set $D$ as
$$\psi_i=C\sum_{S \subseteq D-\{i\}}\frac{\mathcal{U}(S\cup\{i\})-\mathcal{U}(S)}{\binom{n-1}{|S|}}$$ where $C$ is an arbitrary (positive) constant, and $\mathcal{U}(S)$ is the performance score of the model trained on data $S \subseteq D$. One challenge is that computing the exact data Shapley values on large data sets for sophisticated models, such as deep neural networks, is computational prohibitive.  Ghorbani and Zou~\cite{pmlr-v97-ghorbani19c} also develop Monte Carlo and gradient-based methods for estimation.

If a data point $p$ appears in two samples $D_1$ and $D_2$ from the same data distribution, intuitively the Shapley value of $p$ in $D_1$ and $D_2$ should be similar.  Mathematically, the intrinsic Shapley value of $p$ in a distribution should be the expectation of the Shapley value of $p$ in the distribution.  Based on this intuition, Ghorbani~\textit{et al.}~\cite{GKZ20} propose the notion of distributional Shapley.  Let $\mathcal{Z}$ be a universe in question. For example, in classification problems, conventionally $\mathcal{Z}=\mathcal{X} \times \mathcal{Y}$, where $\mathcal{X}$ is the feature space and $\mathcal{Y}$ is the output.  Let $\mathcal{D}$ be a data distribution in $\mathcal{Z}$. Assuming a potential function or a performance metric $U: \mathcal{Z}^* \rightarrow [0, 1]$ and a sample size $m > 0$, the distributional Shapley value of a point $z \in \mathcal{Z}$ is the expected Shapley value over data sets of size $m$ containing $z$, that is, $\nu(z; U, \mathcal{D}, m)=  \mathbb{E}_{S \sim \mathcal{D}^{m-1}}[\psi (z; U, S \cup \{z\}]$, where $S \sim \mathcal{D}^{m-1}$ is a set of $m$ points sampled i.i.d. from $\mathcal{D}$.  They show that distribution Shapley values are stable. Kwon~\textit{et~al.}~\cite{Kwon2020EfficientCA} further derive the computationally tractable expressions for distributional Shapley for a series of models, including linear regression, binary classification and non-parametric density estimation.

Alternative to Shapley values, there are some other data valuation methods.  For example, in machine learning, influence functions~\cite{10.5555/3305381.3305576, DBLP:journals/corr/abs-1912-01321} approximate leave-one-out to assess the value of a data item.  Cai~\textit{et~al.}~\cite{pmlr-v40-Cai15} propose strategy-proof mechanisms for data elicitation and trade off between model accuracy and reward. Richardson~\textit{et~al.}~\cite{richardson2019rewarding} focus on the case of linear regression. Recently, Yoon~\textit{et~al.}~\cite{49189} propose data valuation using reinforcement learning.  They use a data value estimator to learn how much a data item as an element in the training data contributes to improving model performance. One distinct advantage is that the model being trained and the data value estimator can improve each other's performance.

Data quality is an important issue~\cite{10.1145/505248.506010}.  There are many studies on assessment of data quality~\cite{10.1145/505248.506010, doi:10.1080/07421222.1996.11518099, HPKBD15}.  Some studies specifically focus on pricing based on data quality and the impact on data markets.  Heckman~\textit{et~al.}~\cite{HPKBD15} propose a simple linear model, 
$$\textit{Value of data}=\textit{fixed cost} + \sum_{i} w_i \cdot \textit{factor}_i,$$
where the factors include but are not limited to age of data, periodicity of data, volume of data, and accuracy of data, and $w_i$ is the associated weight. One practical difficulty in using the model is that the parameters in the model are hard to estimate.  Another difficulty is that many data sets do not have public prices associated. Yu and Zhang~\cite{YU20171} consider pricing multiple versions formed by multiple factors of data quality and build a two-level model.  The first level is the data platform where a single owner is assumed, who designs the number of versions. The second level is the customers who want to maximize the data utility.  Each level is modeled as a maximization problem and thus the whole model is a bi-level programming problem, which is NP-hard.

Another way to form multiple versions of data products is to charge by queries~\cite{10.1145/2463676.2465335, 10.1145/2213556.2213582, 10.1145/2770870, 10.14778/2367502.2367548}. Intuitively, a data seller may treat a view of a data set as a version.  Setting the price for every possible view is not only tedious but also tricky.  If prices on views are not set properly, arbitrages or less than highest prices may happen.  Koutris \textit{et~al.}~\cite{10.1145/2770870,10.1145/2213556.2213582} propose a framework of query and view based data pricing.  The major idea is that a seller only needs to specify the prices on a few views, and then the prices of other views can be decided algorithmically. Their advocate two desiderata, arbitrage-freeness and discount-freeness.  Theoretically, they show the existence and uniqueness of pricing functions satisfying the requirements.  They also show the complexity of computing the pricing functions.  Unfortunately, only selection views and conjunctive queries without self-joins are tractable. They present polynomial time algorithms for chain queries and cyclic queries.

Technically, the core idea in the view and query based pricing framework is query determinacy~\cite{10.1145/1806907.1806913, 10.1007/11965893_5, 10.1145/1065167.1065174}. A query $Q$ is said to be determined by a set of views $V$ if the answer to $Q$ can be completely derived from the views.  Query determinacy enables the feasibility of arbitrage detection. If $V$ determines $Q$, then arbitrage happens if and only if the price of $V$ is cheaper than that of $Q$.

Koutris~\textit{et~al.}~\cite{10.1145/2463676.2465335} further explore the technical challenges in practical implementation of view and query based data pricing.  Specifically, they develop an integer linear programming formulation for the pricing problem with a large number of queries. Considering the scenario where a user may purchase multiple queries over time or the database is updated, such that information in multiple queries and updates may have overlaps, they also leverage query history to avoid double charging. To handle the situation where there are multiple sellers, they define the share of a seller as the maximum revenue that the seller can get among all minimum-cost solutions, and accordingly define a fair revenue distribution policy.  A prototype demonstration system is reported in~\cite{10.14778/2367502.2367548}.

Tang~\textit{et~al.}~\cite{10.1007/978-3-642-40173-2_31} follow the view and query based pricing framework and consider the minimum granularity of data, that is, each tuple is a view. Their model assigns to each tuple a price and prices queries based on minimal provenances. Tang~\textit{et~al.}~\cite{10.1007/978-3-319-10073-9_3} extend view and query based pricing to XML documents and consider the situation where a customer may just want to purchase a sample instead of the complete query result.

\subsection{Arbitrage-free Pricing}\label{sec:arbitrage-free-pricing}

Arbitrage is probably the most intensively studied issue in pricing data products.  As introduced in Section~\ref{sec:arbitrage-definition}, in general, arbitrage is the activities that take advantage of price differences between two or more markets or channels.  Arbitrage is undesirable in many pricing models.  Unfortunately, arbitrage may sneak in pricing models without rigorous design.  For example, Balazinska~\textit{et~al.}~\cite{journals/pvldb/BalazinskaHS11} analyze that subscription based pricing possibly with a query limit allows arbitrage.  Muschalle~\textit{et~al.}~\cite{10.1007/978-3-642-39872-8_10} point out that a pricing model charging users a certain amount of API calls for a fixed rate may potentially allow arbitrage, depending on the package size.

Arbitrage-freeness is one of the fundamental properties of pricing models in query and view based pricing~\cite{10.1145/2463676.2465335, 10.1145/2213556.2213582, 10.1145/2770870, 10.14778/2367502.2367548}. Li and Miklau~\cite{DBLP:conf/webdb/LiM12} and Li~\textit{et~al.}~\cite{10.1145/2691190.2691191} develop frameworks of pricing linear aggregate queries.  Specifically, Li~\textit{et~al.}~\cite{10.1145/2691190.2691191} consider linear queries. Given a data set of $n$ tuples $x_1, \ldots, x_n$, a linear query $\mathbf{q}=(q_1, \ldots, q_N)$ is a real-valued vector, and the answer $\mathbf{q(x)}=\sum_{i=1}^n q_i x_i$. For a multiset of queries $\mathbf{S=\{Q_1, \ldots, Q_k\}}$ and query $\mathbf{Q}$, if the answer to $\mathbf{Q}$ can be linearly derived from the answers to the queries in $\mathbf{S}$, then $\mathbf{Q}$ is said to be determined by $\mathbf{S}$, denoted by $\mathbf{S \rightarrow Q}$. A pricing function $\pi(\mathbf{Q})$ is arbitrage-free if for any multiset $\mathbf{S}$ and query $\mathbf{Q}$ such that $\mathbf{S \rightarrow Q}$, $\pi(\mathbf{Q}) \leq \sum_{i=1}^k \pi(\mathbf{Q}_i)$.

Under the general intuition of arbitrage-freeness, Li~\textit{et~al.}~\cite{10.1145/2691190.2691191} consider a specific form of queries, linear queries with variance $(\mathbf{q}, v)$, that is, the estimation of the answer to query $\mathbf{q}$ should have a variance no larger than $v$. Using different values of $v$, different versions are formed.  A pricing model not carefully designed may allow arbitrage. 

Li~\textit{et~al.}~\cite{10.1145/2691190.2691191} first establish the observation that $\pi(\mathbf{q}, v) = \Omega(\frac 1 v)$. Then, they synthesize pricing function $\pi(\mathbf{q}, v)=\frac{f^2(\mathbf{q})}v$, which is arbitrage-free if $f$ is positive and semi-norm\footnote{A function $f: \mathcal{R}^n \rightarrow \mathcal{R}$ is semi-norm if for any $c \in \mathcal{R}$ and any query $\mathbf{Q} \in \mathcal{R}^n$, $f(c\mathbf{q})=|c|f(\mathbf{q})$; and for any $\mathbf{q}_1, \mathbf{q}_2 \in \mathcal{R}^n$, $f(\mathbf{q}_1+\mathbf{q}_2) \leq f(\mathbf{q}_1)+f(\mathbf{q}_2)$.}.
For any arbitrage-free pricing functions $\pi_1, \ldots, \pi_k$, $f(\pi_1(\mathbf{q}), \ldots, \pi_k(\mathbf{q}))$ is also arbitrage-free if $f$ is a subadditive\footnote{A function $f$ is subadditive if for any $x_1, \ldots, x_k$, $f(\sum_{i=1}^k x_i) \leq \sum_{i=1}^k f(x_i)$.}  and nondecreasing function.

As Roth~\cite{10.1145/3139455} summarizes, the framework by Li~\textit{et~al.}~\cite{10.1145/2691190.2691191} still faces three important challenges.  First, arbitrage is still possible to derive answers to a bundle of queries from another bundle of queries and their answers.  Second, arbitrage is still possible on biased estimators for statistical queries.  Last, it is unclear whether we can obtain arbitrage-free pricing maximizing profit given the distribution of buyer demands. Later, Deep and Koutris~\cite{deep2016design} provide some interesting insights to arbitrage-free pricing for bundles.

Lin and Kifer~\cite{10.14778/2732939.2732948} investigate arbitrage-free pricing for general data queries.  They consider three types of pricing models for query bundles, where a query bundle is a set of queries posted simultaneously as a batch.  First, an instance-independent pricing function depends on the query bundle but not the database instance.  Second, an up-front dependent pricing function depends on both the query bundle and the database instance. A customer knows an un-front dependent pricing function, and decides whether to purchase or not the query answers. Last, a delayed pricing function depends on both the query bundle and the answers computed by the query bundle on the current database instance.  The customer knows the pricing function, but do not know the exact price.  Once agreeing, the customer is charged when the answers are computed.

Lin and Kifer~\cite{10.14778/2732939.2732948} also summarize five different types of arbitrage situations.  First, if prices are quoted by queries, in order to avoid price-based arbitrage, answers to queries should not be deduced from prices along.  Second, a buyer may use multiple accounts to derive answers to a query bundle.  To avoid separate account arbitrage, the price of a query bundle $[q_1, q_2]$ should be at most the sum of the prices of $q_1$ and $q_2$.   Third, if the answers to a query bundle $q'$ can always be deduced from answers to another query bundle $q$, to prevent post-processing arbitrage from happening, the price of $q$ should be no cheaper than that of $q'$. Fourth, although the answers to a query bundle $q$ may not be always derivable from the answers to another query bundle $q'$ on all database instances, still for a specific database instance $\mathcal{I}$, the answers to $q$ may be derived from the answers to $q'$.  If so, a serendipitous arbitrage happens.  Last, if two queries behave almost identical but their prices are dramatically different, almost-certain arbitrage happens.  Based on the above categorization, they discuss conditions that can prevent various types of arbitrage situations from happening.

Pricing many queries in real time with formal guarantees on arbitrage freeness is challenging. Many theoretical methods are not scalable in practice.  For example, it takes QueryMarket~\cite{10.1145/2463676.2465335} about one minute to compute the price of a join query over a relation of about 1000 tuples. Qirana~\cite{10.1145/3035918.3064017, 10.14778/3137765.3137816} is a system for query-based pricing.  The system allows data sellers to choose from a set of pricing functions that are information arbitrage-free, which covers both post-processing arbitrage-freeness and serendipitous arbitrage-freeness in Lin and Kifer's taxonomy~\cite{10.14778/2732939.2732948}. Qirana also supports history-aware pricing.  Qirana has been shown highly efficient and scalable on TPC-H\footnote{\url{http://www.tpc.org/tpch.}} and SSB\footnote{\url{http://www.cs.umb.edu/?poneil/StarSchemaB.PDF}} benchmark datasets as demonstration. 

The key idea in Qirana is that it regards a query as an uncertainty reduction mechanism.  Initially, a buyer faces a set of possible databases $\mathcal{I}$ defined by a database schema, primary keys and predefined constraints.  Once a buyer obtains the answer $E$ to a query $Q$, all possible databases $D$ such that $E \neq Q(D)$ are eliminated.  The price assigned to $Q$ should be a function of how much the set of possible databases shrinks.  Let $\mathcal{S}$ be the set of possible databases before the query $Q$ is answered. $\mathcal{S}$ is called the support set.  Then, a weighted coverage function assigns a weight $w_i$ to every $D_i \in \mathcal{S}$, and computes the price to a query by $p^{wc}(Q, D)=\sum_{Q(D_i) \neq Q(D)}w_i$.  Alternatively, consider the equivalence relation in $\mathcal{S}$: $D_i \sim D_j$ if and only if $Q(D_i)=Q(D_j)$.  Assign to each possible database $D_i \in \mathcal{S}$ a weight $w_i$ such that $\sum_{D_i \in \mathcal{S}}w_i=1$.  Let $\mathcal{P}_Q$ be the set of equivalence classes.  For each class $B \in \mathcal{P}_Q$, denote by $w_B=\sum_{D_i \in B}w_i$.  The Shannon entropy function is used to compute the price of query $Q$ as the entropy of the query output $P^H(Q, D)= - \sum_{B \in \mathcal{P}_Q}w_B \log w_B$. The q-entropy function (also known as Tsallis entropy) for $q=2$ is used to assign to $Q$ the price $P^T(Q, D) = \sum_{B \in \mathcal{P}_Q} w_B(1-w_B)$. Deep and Koutris~\cite{deep2016design} show that the weighted coverage function, the Shannon entropy function and the 2-entropy function are all arbitrage-free.  

Using the complete set of possible databases as the support set leads to a $\#P$-hard problem. To make the price calculation computationally feasible, Qirana uses uniform random sample and random neighboors as the support sets.

In targeted advertising markets, user data, such as opt-in email addresses, and user impressions are sold as data products. How to price users\footnote{Here, ``buying a user'' is short for purchasing the impression of a user in online advertising and a user email in targeted email advertising, for example.} properly to avoid arbitrage is important.  Xia and Muthukrishnan~\cite{10.5555/3237383.3237436} consider the following problem. Denote by $q_i$ a selection query over user attributes, by $U_i$ the set of all users satisfying $q_i$, and by $p_i$ the price of each user in $U_i$.  If a buyer purchases $n$ users ($1 \leq n \leq |U_i|$) in $U_i$, she/he has to pay $n\cdot p_i$. If prices of different queries are not well coordinated, version-arbitrage may arise. If two queries $q_i$ and $q_j$ return similar user sets but $q_i$ is dramatically more expensive than $q_j$, then a user who wants $q_i$ may purchase $q_j$ instead. Xia and Muthukrishnan~\cite{10.5555/3237383.3237436} point out that uniform pricing, that is, every query has the same price, is arbitrage-free, but is a logarithmic approximation to the maximum revenue arbitrage-free pricing solution.  Then, they present a greedy non-uniform pricing design. The design starts with the optimal uniform pricing that is arbitrage-free, and then iteratively updates the pricing function.  If the price of a query can be updated to increase the revenue, it is increased so that the arbitrage-free property is retained.  This greedy algorithm is still a logarithmic approximation to the maximum revenue arbitrage-free pricing solution.

Chen~\textit{et~al.}~\cite{10.1145/3299869.3300078} develop an arbitrage-free pricing design for multiple versions of a machine learning model. They assume that a broker trains the optimal model on the complete raw data.  Then, random Gaussian noises are added to the optimal model to produce different versions for different buyers. The assumption is that the error of a machine learning model instance is monotonic with respect to the variance of the noise injected into the model. In this setting, a pricing function is arbitrage-free if and only if the price of a randomized model instance is monotonically increasing and subadditive with respect to the inverse of the variance.

\subsection{Revenue Maximization Pricing}

As explained in Section~\ref{sec:revenuemax}, the objective of revenue maximization is often of special interest in designing pricing strategies, since for a business to be successful long term, a more immediate and important requirement is to win over as many customers as possible. 

Revenue maximization pricing for data products is a relatively less explored area.  A possible reason is that, comparing with pricing digital products, some other factors in pricing data products need more urgent accommodation, such as arbitrage.

As mentioned in Section~\ref{sec:arbitrage-free-pricing}, Xia and Muthukrishnan~\cite{10.5555/3237383.3237436} develop logarithmic approximation pricing algorithms for revenue maximization in user-based markets. They also consider the situations where both the maximum number (i.e., maximum demand) and the minimum number (i.e., minimum demand) of users that a buyer purchases are specified, and provide an $O(D)$ approximation algorithm to maximize revenue, where $D$ is the largest minimal demand among all buyers. 

Chawla~\textit{et~al.}~\cite{10.14778/3357377.3357378} consider query and view based pricing for arbitrage-free revenue maximization under the assumption that all buyers are single-minded and the supply is unlimited. A buyer is single-minded if the buyer wants to purchase the answer to a single set of queries.  They consider three types of pricing functions.  Uniform bundle pricing sets the price of every bundle identical.  Additive or item pricing prices each item and charges a bundle the sum of prices for the items in the bundle.  Fractionally subadditive pricing or XOS sets $k$ weights $w_j^1, \ldots, w_j^k$ for each item $j$, and for a bundle $e$, the price is set to $\max_{i=1}^k \sum_{j \in e}w_j^i$. Building on the extensive studies on revenue maximization with single-minded buyers and unlimited supply~\cite{10.1145/1134707.1134711, 10.5555/1109557.1109678, 10.5555/1070432.1070598}, they develop new heuristics. 

It is well known that there exists uniform bundle pricing that is $O(\log m)$ approximation of revenue maximization, where $m$ is the number of bundles.  Swamy and Cheung~\cite{SwamyCheung08} show that item pricing can achieve an $O(\log B)$ approximation of maximum revenue, where $B$ is the maximum number of bundles an item can involve. Chawla~\textit{et~al.}~\cite{10.14778/3357377.3357378} show some new lower bounds, that is, uniform bundle pricing, item pricing and XOS pricing combining a constant number of item pricing functions are still $\Omega(\log m)$ away from maximum revenue.  They also present approximation algorithms.

To maximize revenue in machine learning models, Chen~\textit{et~al.}~\cite{10.1145/3299869.3300078} show that the optimization problem is coNP-hard. Thus, they relax the subadditive constraint $p(x+y) \leq p(x) + p(y)$ by $\frac{q(x)}{x} \geq \frac{q(y)}{y}$ for every $0 < x \leq y$, and turn to finding a pricing function $q()$ such that $\frac{q(x)}{x}$ is decreasing with respect to $x$. They show that, for every well standing pricing function $p()$, there exists a pricing function $q()$ with the relaxed subadditive constraint such that $\frac{p(x)}{2} \leq q(x) \leq p(x)$, and $q(x)$ can be computed using dynamic programming in $O(n^2)$ time, where $n$ is the number of interpolated price points.

\subsection{Fair and Truthful Pricing }\label{sec:fair-pricing}

Fairness and truthfulness are important for data product markets. Recall that fairness refers to that the revenue generated by a sale transaction in the data market is distributed among sellers in an unprejudiced manner so that they are paid for their marginal contributions.  Truthfulness means a market where buyers are well motivated to report their internal valuations of data products unwarily.

Agarwal~\textit{et~al.}~\cite{10.1145/3328526.3329589} propose a mathematical model of data marketplaces that are fair, truthful, revenue maximizing, and scalable.  They assume each seller $j$ supplies a data stream $X_j$ and each buyer $n$ conducts a prediction task $Y_n$, where $X_j, Y_n \in \mathcal{R}^T$.  For example, $X_j$ may be a stream of customers' interest on different products, and $Y_n$ is a task predicting a new customer's interest.  Taking a prediction task $Y_n$ and an estimate $\hat{Y}_n$, a prediction gain function $\mathcal{G}_n: \mathcal{R}^{2T} \rightarrow [0, 1]$ measures the quality of the prediction.  The value that buyer $n$ gets from estimate $\hat{Y}_n$ is $\mu_n\cdot \mathcal{G}(Y_n, \hat{Y}_n)$, where $\mu_n$ is the price rate that the buyer is willing to pay for a unit increase in $\mathcal{G}$.  A machine learning model $\mathcal{M}: \mathcal{R}^{MT} \rightarrow \mathcal{R}^T$ uses data from $M$ sellers to produce an estimate $\mathcal{Y}_n$ for buyer $n$'s prediction task $Y_n$. Let $p_n$ and $b_n$ be the price and the bid, respectively. Then, allocation function $\mathcal{AF}: (p_n, b_n; X_M) \rightarrow \widetilde{X}_M$ measures the quality at which buyer $n$ obtains that is allocated to the sellers on sale $X_M$, where $\widetilde{X}_M \in \mathcal{R}^M$.  Revenue function $\mathcal{RF}: (p_n, b_n, Y_n; \mathcal{M}, \mathcal{G}, X_M) \rightarrow r_n$ calculates how much revenue $r_n \in \mathcal{R}^+$ to extract from the buyer.  The utility that buyer $n$ receives by bidding $n_n$ for $Y_n$ is $$\mathcal{U}(b_n, Y_n) = \mu_n \cdot \mathcal{G}(Y_n, \hat{Y}_n)-\mathcal{RF}(p_n, b_n, Y_n),$$ where $\hat{Y}_n = \mathcal{M}(Y_n, \widetilde{X}_M)$ and $\widetilde{X}_M=\mathcal{AF}(p_n, b_n; X_M)$.  A market is truthful if for all prediction tasks $Y_n$, $\mu_n = \arg\max_{z \in \mathcal{R}^+} \mathcal{U}(z, Y_n)$.  They adopt the notion of fairness following the famous Shapley fairness~\cite{Shapley}.

One main result~\cite{10.1145/3328526.3329589} is that, the data market defined as such is truthful if and only if function $\mathcal{AF}*$ is monotonic, that is, an increase in the difference between price rate $p_n$ and bid $b_n$ leads to a decrease in predication gain $\mathcal{G}$.  They also give randomized $\epsilon$-approximation algorithms for fair data market, that is, $||\psi_{n, \text{Shapley}}-\hat{\psi}_n||_\infty < \epsilon$ with probability $1-\delta$, where $\psi_{n, \text{Shapley}}$ is the Shapley-fair payment division among sellers, $\hat{\psi}_n$ is the output of the approximation algorithm, and $\delta, \epsilon >0$. Their algorithms are polynomial. 

Shapley fairness~\cite{Shapley} is popularly adopted as the foundation of fairness in data markets. However, computing Shapley value is exponential~\cite{doi:10.1287/moor.19.2.257}.  Maleki~\textit{et~al.}~\cite{maleki2013bounding} present a permutation sampling method that approximates Shapley values for any bounded utility functions. The basic idea is to use Equation~\ref{eq:shapley-permutation} and tackle $\psi(s)=E[\mathcal{U}(P_s^\pi\cup\{s\})-\mathcal{U}(P_i^\pi)]$ by sample mean. Following Hoeffding's inequality~\cite{10.2307/2282952}, to achieve an $(\epsilon, \delta)$-approximation, that is, $P(|\hat{s}-s|_p \leq \epsilon) \geq 1 - \delta$, where $\hat{s}$ is the estimate, we need $\frac{2r^2N}{\epsilon^2}\log\frac{2N}\delta$ samples and evaluate the utility function $O(N^2 \log N)$ times, where $r$ is the range of the utility function $\mathcal{U}$. 

Jia~\textit{et~al.}~\cite{pmlr-v89-jia19a} present approximation algorithms for Shapley value that can substantially reduce the number of times that the utility function is evaluated.  First, they apply the idea of feature selection using group testing~\cite{10.5555/2969033.2969223, doi:10.1142/4252}.  For user $s$, let $\beta_s$ be the random variable that $s$ appears in a random sample of sellers.  Then, for sellers $s_i$ and $s_j$, the difference in Shapley values between $s_i$ and $s_j$ is $$
\begin{array}{rcl}
\psi(s_i)-\psi(s_j)&=&\frac 1 {N-1}\sum_{S \in D \setminus\{s_i, s_j\}} \frac{\mathcal{U}(S\cup\{s_i\})-\mathcal{U}(S\cup\{s_j\})}{\binom{N-2} {|S|}}\\
&=&E[(\beta_{s_i}-\beta_{s_j})\mathcal{U}(\beta_{s_1}, \ldots, \beta_{s_j})]
\end{array}
$$
where $\mathcal{U}(\beta_{s_1}, \ldots, \beta_{s_j})$ is the utility computed using the sellers appearing in the random sample. They can use group testing to first estimate the Shapley differences and then derive the Shapley value from the differences by solving a feasibility problem.  They show that this algorithm is an $(\epsilon, \delta)$-approximation that evaluates the utility function at most $O(\sqrt{N}(\log N)^2)$ times.  They further observe that most of the Shapley values are around the mean.  Exploiting this approximate sparsity, they give an $(\epsilon, \delta)$-approximation algorithm that evaluates the utility function only $O(N(\log N) \log(\log N)$ times.

Ghorbani and Zou~\cite{pmlr-v97-ghorbani19c} propose a principled framework of fair data evaluation in supervised learning, and Monte-Carlo and gradient-based approximation methods.  Their Monte-Carlo method follows a general idea similar to that in Jia~\textit{et~al.}~\cite{pmlr-v89-jia19a}. They generate Monte-Carlo estimates until the average empirically converges. They also argue that, in practice, it is sufficient to estimate Shapley values up to the intrinsic noise in the predictive performance on the test data set. Adding one tuple as a training data point does not significantly affect the performance of a model trained using a large training data set.  Therefore, truncation can be used in practice based on the bootstrap variation on the test set.  In their gradient Shapley method, they train a model using one ``epoch'' of the training data, and then update the model by gradient descent on one data point at a time, where the marginal contribution is the change in the performance of the model.

In general, computing Shapley values requires an exponential number of model evaluations.  However, for some specific model, the computation may be reduced dramatically.  For example, Jia~\textit{et~al.}~\cite{10.14778/3342263.3342637} show that for unweighted kNN classifiers, the exact computation needs only $O(N \log N)$ time and an $(\epsilon, \delta)$-approximation can be achieved in $O(N^{h(\epsilon, k)}\log N)$ time when $\epsilon$ is not too small and $k$ is not too large.  They also propose a Monte-Carlo approximation of $O(\frac{N(\log N)^2}{(\log k)^2})$ for weighted kNN classifiers. A key enabler of the progress is the specific utility function of a kNN classifier $$\mathcal{U}_{kNN}(S)=\frac 1 k \sum_{i=1}^{\min\{k, |S|\}}\mathds{1}[y_{\alpha_i(S)}=y_{\text{test}}]$$ where $\alpha_i(S)$ is the index of the training feature that is the $k$-th closest to $x_\text{test}$ among the training examples in $S$.  Moreover, the sublinear approximation for unweighted kNN classifiers is facilitated by locality sensitive hashing~\cite{10.1145/997817.997857}. 

Recently, Jia~\textit{et~al.}~\cite{DBLP:journals/corr/abs-1911-07128} leverage the efficient computation of Shapley values in kNN~\cite{10.14778/3342263.3342637} to tackle general classification problems.  They propose to first train a target model, such as a deep neural network, and identify the features.  Then, they conduct a model distillation to kNN by training a kNN classifier using the features to mimic the performance of the original model and tune parameter $k$, the number of nearest neighbors considered.  Last, they apply the Shapley value estimation method in kNN~\cite{10.14778/3342263.3342637} to approach the Shapley values in the target model.

Many classic rewarding methods, such as Shapley values, may be vulnerable to data-replication attacks.  One data provider may replicate its data and act as an additional provider to obtain extra unconscionable rewards. To prevent data-replication attacks from happening, replication-robust payoff mechanisms are proposed.  Han~\textit{et~al.}~\cite{Han2020ReplicationRobustPW} propose a fix to Shapley value based payoff mechanisms. The idea is to down-weigh the Shapley value -- a data provider gets a less reward if there are multiple copies of its data in the coalitions. 


Related to fairness and truthfulness in a market, cooperation among different agents in a market may happen. Building trust in a sub-community within a data marketplace becomes an interesting subject. Armstrong and Durfee~\cite{10.5555/551984.852255} analyze factors that may influence the efficiency of building trust and conducting cooperation in a data market.  For each agent in a market, the other agents can be divided into two categories, namely those remembered agents and those strange or forgotten agents. They have a few interesting findings.  Cooperations arising from iterated interactions is inversely proportional to the rate of system mixing,  the number of initially misbehaving agents, and the rate at which agents explore alternative strategies. Cooperation is also initially inversely proportional to population size. At the same time, cooperation is proportional to average member size and better estimation of the likelihood of strange agents to misbehave.


\subsection{Privacy Preserving Marketplaces of Data}

Privacy is a serious concern and also a critical tipping point in designing marketplaces of data.  When a user shares her/his data with some others, the user may disclose her/his privacy to some extent.  Therefore, it is important to explore how to protect or minimize the privacy leakage.  At the same time, it is also important to understand how a seller's privacy disclosure may be properly compensated through data pricing.

Ghosh and Roth~\cite{10.1145/1993574.1993605} design truthful marketplaces where data buyers want to purchase data to estimate statistics and sellers want compensation for their privacy loss. In the design, there is only one query and the individual evaluations of their data are private. Data owners are asked to report the costs for the use of their data. Under the assumption of differential privacy~\cite{10.1007/11681878_14, 10.1007/978-3-540-79228-4_1}, they transform the problem into variants of multi-unit procurement auction.  They show that, when a buyer holds an accuracy goal, the classic Vickrey auction can minimize the buyer's total cost and guarantee the accuracy.  When the buyer has a budget, they give an approximation algorithm to maximize the accuracy under the budget constraint.

The method by Ghosh and Roth~\cite{10.1145/1993574.1993605} may not work well when the costs and the data are correlated.  For example, a store with more customer traffic may request a higher cost in using the data. Correspondingly, reporting the cost may reveal the privacy of the store. Fleischer and Lyu~\cite{10.1145/2229012.2229054} tackle the scenario where costs are correlated with data and propose a posted-price-like mechanism.  Given a set of data sellers categorized into different types and the associated distributions of costs, the mechanism offers each user a contract with the expected payment corresponding to the type. If a seller takes the offer, the payment is determined by the seller's verifiable type and the associated payment in the contract.  All sellers have the same probability to take or reject their contracts independently.  The sellers are truthful, that is, a user takes the offer if the payment is larger than or equal to the privacy loss.  This posted-price-like mechanism is Bayesian incentive compatible (i.e., every seller's strategy is Bayesian-Nash equilibrium), ex-interim individually rational (i.e., the expected utility is non-negative for every seller when the seller decides truthfully), $O(\epsilon^{-1})$-accurate, perfectly data private (i.e., whenever the mechanism's posterior belief about a seller's data differs from its prior belief, the mechanism pays the seller) and $\epsilon$-differentially private.

Li~\textit{et~al.}~\cite{10.1145/2691190.2691191} tackle the same problem as Ghosh and Roth~\cite{10.1145/1993574.1993605} do, but assume that individual valuations are public and focus on returning unbiased estimations and pricing multiple queries consistently.  To address the concerns on privacy loss, they develop a theoretical framework to divide the price among data owners who contribute to the aggregate computation and thus have loss of privacy. Their framework extends several principles from both differential privacy and query pricing in data markets.

The fairness mechanism considered by Li~\textit{et~al.}~\cite{10.1145/2691190.2691191} only compensates a seller whose data are used.  Niu~\textit{et~al.}~\cite{10.1145/3219819.3220013} further consider the scenario where multiple sellers' data are correlated and extend to dependent fairness. In dependent fairness, a seller $s$ is still compensated if the data of another seller $s'$ are used that are correlated with the data of $s$.  They propose two approaches to privacy compensation.  In the bottom-up approach, the broker first satisfies each individual seller's privacy compensation and then decides the price for the statistic selling to a buyer.  In the top-down design, the broker decides the total price of a data aggregate product sold to a buyer, and then spares a fraction of the total price for privacy compensation.  The privacy compensation is divided and assigned to individual data sellers by solving a budget allocation problem.  Each seller receives a compensation roughly proportional to the privacy loss due to the data sharing.  Niu~\textit{et~al.}~\cite{8737579} further extend to time series data that may have temporal correlations.  They adopt Pufferfish privacy~\cite{10.1145/2213556.2213571} to measure privacy losses under temporal correlations.

While various efforts have been made to address the challenges of privacy loss compensation when user data are correlated in one way or another, as Ghosh and Roth~\cite{10.1145/1993574.1993605} point out, in general, it is impossible for any mechanism to compensate individuals for privacy loss properly if correlations between their private data and their cost functions are unknown beforehand.

In the classical setting of physical goods~\cite{10.2307/2555674}, using contract theory~\cite{RePEc:oxp:obooks:9780195102680} with hidden information, that is, unobservable types of buyers, a seller can design a set of contracts with different consumption levels to maximize revenue from buyers. Naghizadeh and Sinha~\cite{10.1145/3328526.3329633} extend the contract design model to price a bundle of queries at different privacy levels to maximize revenue.  They also consider adversarial users.  Their work also adopts differential privacy~\cite{10.1007/11681878_14, 10.1007/978-3-540-79228-4_1}. For a query bundle $\{Q_1, \ldots, Q_k\}$, a contract is a tuple $(p, \epsilon, s)$, where $p>0$ is the price paid by a buyer, $\epsilon$ is the privacy budget, such that a buyer can get an answer to query $Q_i$ $(1 \leq i \leq k)$ with $\epsilon_i$-differential privacy guarantee, and $\epsilon \geq \sum_{i=1}^k \epsilon_i$, and $p$ is the post-hoc fine to be paid if the buyer is found misusing the query answers.  It is assumed that an adversarial buyer derives a benefit $C(\epsilon)$, which is monotonically increasing and convex, $C(0)=0$.  One interesting finding is that, in the traditional contract theory, if there are $n$ types of honest buyers and one type of adversarial buyers, the seller should design up to $n+1$ contracts.  In the data marketplace situation, they show that up to $n$ contracts are sufficient.  In other words, a data seller should not design a contract for the adversary. Instead, the seller should adjust the contracts' pricing to account for the risks from adversarial users. They also design post-hoc fines in pricing query bundles that can help to reduce loss due to privacy leakage by adversarial buyers.  They provide a fast approximation algorithm to compute the contracts.

A data owner has to decide a tradeoff between privacy and data utility.  Li and Raghunathan~\cite{LI201463} design an economics-based incentive-compatible mechanism for a data owner to price and disseminate private data.  Specifically, let two-part tariff pricing function $R(s, x)=\alpha_s+\beta_sx$ be the price for $x$ amount of data at sensitivity level $s$, where $\alpha_s$ and $\beta_s$ are the fixed and variable price factors, respectively. Assuming two types of data users, one type for aggregate information and patterns in data and the other type for individual identity and personal information, the proposed mechanism works in four stages.  First, the data owner selects a variety of sensitivity types to offer. Second, the data owner offers different prices for data with different sensitivity types. Third, a data user selects a certain sensitivity type with corresponding price, and thus reveals the user type.  Last, the data user selects the optimal amount of data with the chosen sensitivity type. The core idea is that the data owner can identify the sensitive attributes in the data, such as the identifying attributes, which are not useful for aggregate analysis but necessary at individual communication.  A data owner can offer a lower price for data without sensitive attributes, and charge for a higher price for data with sensitive attributes.  This approach provides an orthogonal idea to the popular ways of tuning the parameter in differential privacy.

Due to the privacy concerns, when a company may have opportunities to collect data about its customers, should it do it (i.e., collecting and revealing the data) or not (i.e., a blanket policy of never collecting)?  Jaisingh \textit{et~al.}~\cite{JAISINGH2008857} find that the company should not collect customer data if the total gains from trading the data cannot cover the privacy loss. In practice, there is an increasing tendency for consumers to overestimate their loss of privacy, particularly when the use of the private data is uncertain.  In other cases, the company should offer two contracts on their services and products. One contract collects the customer data at a certain price, and the other contract does not collect any customer data at a different price.

While most of the studies on privacy preserving data marketplaces focus on the privacy of data owners, transactions may also disclose privacy of data buyers, such as what, when and how much they buy.  For example, a retail company purchasing query results may consider what queries (e.g., the products or customer groups involved in the queries), when (e.g., the periods where the queries are concerned), and how much data it purchases as privacy, and may want to keep the information confidential from any others, including the data sellers and the broker.  Aiello~\textit{et~al.}~\cite{eurocrypt-2001-2009} design a mechanism such that after making an initial deposit and maintaining a sufficient balance, a buyer can engage in an unlimited number of price-oblivious transfer protocols where the sellers and the broker cannot know anything other than the amount of interaction and the initial deposit amount.  The broker even cannot know the buyer's current balance and when the buyer's balance runs out.  This is achieved by adapting conditional disclosure~\cite{10.1006/jcss.1999.1689} to the two-party setting.

Distribution and use of private data are another important step where privacy may leak. Hynes~\textit{et~al.}~\cite{10.14778/3229863.3236266} demonstrate Sterling, a  decentralized marketplace for private data, which supports privacy-preserving distribution and use of data.  The central technical idea comes from privacy-preserving smart contracts on a permissionless blockchain.  To provide strong security and privacy guarantees, they combine blockchain smart contracts, trusted execution environments and differential privacy.  Particularly, smart contracts allow enforcement of constraints on data usage and enables payments and rewards.

\subsection{Data Pricing in Novel Applications: Dynamic Data Pricing, Online Pricing and Federated Learning Pricing}

The demand of data pricing arises in many novel application scenarios. In this subsection, we particularly discuss three emerging situations: dynamic data pricing, online pricing and pricing in federated learning.

Many applications are built on dynamic and online data.  How to price temporal views on data streams properly is an important issue for practical data markets. One central task is to estimate and optimize the operational costs, which are the costs to evaluate queries of different users on the fly.  The pricing decisions involve not only data sellers but also data buyers.  For example, suppose two data buyers $b_1$ and $b_2$ purchase two queries $q_1$ and $q_2$, such that $q_2$ can be written as a further selection on top of $q_1$ (e.g., $q_1$ is about all customers in North America, while $q_2$ keeps all the same as $q_1$ but focuses on only customers in Canada). The optimal pricing of $q_1$ and $q_2$ should take the advantage of the overlap between the two queries so that the sharing can save the operational costs, and, at the same time, be fair to $b_1$ and $b_2$.

Al-Kiswany~\textit{et~al.}~\cite{10.1145/2452376.2452447} propose a greedy method that enumerates all possible sharing plans and selects the one with the minimum additional cost.  It does not come with any quality guarantee. Liu and Hacig\"{u}m\"{u}\c{s}~\cite{10.1145/2588555.2593679} propose an improved method that takes some risk in sharing plan. If the costs of the previous sharings are already cumulated to a high level, and the additional cost of a new sharing (i.e., the risk) is moderate and can be amortized well by the previous sharings, then the new sharing may be taken.  They also give five rules to ensure fair pricing.  Let $AC(S)$ be the cost attributed to a sharing $S$. First, for two identical sharings $S_1=S_2$, $AC(S_1)=AC(S_2)$ should hold.  Second, for any sharing $S$, $AC(S)$ should be no higher than the lowest cost of $S$ if no other sharing exists.  Third, for two sharings $S_1$ and $S_2$, if the query of $S_1$ is contained by the query of $S_2$, that is, the result of $S_1$ is a subset of the result of $S_2$, and the lowest cost of $S_1$ is smaller than the lowest cost of $S_2$ if no other sharing exists, then $AC(S_1) \leq AC(S_2)$. Fourth, a sharing plan with common subexpressions with other sharings should be compensated.  Last, the cost of the global plan should be equal to the sum of costs attributed to all sharings.

In order to purchase dynamic data, a buyer may have to call a seller's API repeatedly. A buyer may have to pay for the same data multiple times. Upadhyaya~\textit{et~al.}~\cite{10.14778/3007328.3007335} explore how to modify APIs to achieve optimal history-aware pricing, that is, buyers are charged only once for data purchased and not updated.  The central idea is the introduction of the notion of refund -- a user can ask for refunds of data that she/he has bought before.  For each query, the seller issues a coupon in addition to the query result, where the coupon records the identity information of the data in the query result.  Specifically, a coupon $c=((id, uid, v), \tau, \mathcal{H}(id \oplus \tau \oplus \kappa))$, where $id$ is a tuple identifier, $uid$ is a user-id, $v$ is a version-id that is monotonically increasing, $\tau$ is a query identifier that is also monotonically increasing, $\mathcal{H}$ is a cryptographic hash function~\cite{10.1109/TIT.1976.1055638}, such as SHA-1, SHA-256 and SHA-3, and $\kappa$ is a secret key only known to the seller.  If a buyer gets two coupons $c_1$ and $c_2$ in two different purchases such that $c_1[(tid, uid, v)]=c_2[(tid, uid, v)]$, then the buyer can ask the seller for a refund by showing the two coupons.  As pointed out by Deep and Koutris~\cite{10.1145/3035918.3064017}, the refund mechanism does not provide any arbitrage-free guarantee.

Qirana~\cite{10.1145/3035918.3064017, 10.14778/3137765.3137816} can support history-aware pricing.  To incorporate a query history, suppose a buyer already purchases queries $\mathbf{Q}=Q_1, \ldots, Q_k$ and pays for a total of $p(\mathbf{Q}, D)$ so far.  When a new query $Q_{k+1}$ comes, let the support set $\mathcal{S}_{k+1}=\{D_i \in \mathcal{S} \mid \mathbf{Q}(D_i)=\mathbf{Q}(D), Q_{k+1}(D_i) \neq Q_{k+1}(D)\}$.  Then, the new total price $p((Q_1, \ldots, Q_k, Q_{k+1}), D)=p(\mathbf{Q},D)+\sum_{D_i \in \mathcal{S}_{k+1}}w_i$.  This history-aware pricing function is shown arbitrage-free.

Zheng~\textit{et~al.}~\cite{10.1145/3084041.3084044} consider online pricing for mobile crowd-sensing data markets.  Different from most of the work on data markets, they assume that data providers are distributed in space and there are three types of spatial queries from buyers, namely single-data query (e.g., inquiring the value at a specific location), multi-data query (e.g., inquiring the mean in a region) and range query (e.g., inquiring the probability that the data at a region falls in a given range).  The vendor uses raw data from data providers and produces a statistical model through Gaussian process to answer queries. To form different versions of data products, the vendor generates different conditional Gaussian distribution with respect to locations and uses the conditional entropy to quantify the quality of the versions.  They propose a randomized online pricing strategy so that the price can be adaptive from the historical queries.  They show that the pricing mechanism is arbitrage-free and is a constant factor approximation of revenue maximization.

Niu~\textit{et~al.}~\cite{Niu2019OnlinePW} consider online data market where a query may be sold to different buyers at different time and the broker can adjust prices over time.  The objective is to maximize the broker's cumulative revenue by posting reasonable prices for sequential queries. They design a contextual dynamic pricing mechanism with the reserve price constraint. The central idea is to use the properties of ellipsoid for efficient online optimization.  Their method can support both linear and non-linear market value models with uncertainty.

Federated learning~\cite{federatedlearning, pmlr-v54-mcmahan17a} trains a machine learning model across multiple decentralized parties, where each party holds local data without any peer-wise data exchanging. The parties and their data sets are often ordered in a federated learning process. To accommodate the participation order and value data in federated learning, Wang~\textit{et~al.}~\cite{wang2020principled} develop federated Shapley value. Let $I$ be the set of participants and $\mathcal{U}$ be the utility function, where $\mathcal{U}(A+B)$ is the utility of training first on $A$ and then on $B$.  For participant $i$ at round $t$ in a federated learning process, the federated Shapley value is 
$$
\psi_t(i)
= \frac 1 {|I_t|}\sum_{S \subseteq I_t\setminus \{i\}} \frac{1}{\binom{|I_t|-1}{|S|}}[\mathcal{U}(I_{1:t-1}+S \cup\{i\}))-\mathcal{U}(I_{1:t-1}+S)]
$$
if $i \in I_t$ and $\psi_t(i)=0$ otherwise. The federated Shapley value of a party is the sum of the values of all rounds, that is, $\psi(i)=\sum_{t=1}^T \psi_t(i)$.  Wang~\textit{et~al.}~\cite{wang2020principled} show that the federated Shapley values have instantaneous group rationality, that is, $\sum_{i \in I_t} \psi_t(i)=\mathcal{U}(I_{1:t})-\mathcal{U}(I_{1:t-1})$.  The fairness is guaranteed at each round.  That is, for any two parties $i$ and $j$, $\psi_t(i)=\psi_t(j)$ at round $t$ if $\forall S \subseteq I_t \setminus \{i, j\}$, $\mathcal{U}(I_{1:t-1}+(S \cup \{i\})) = \mathcal{U}(I_{1: t-1}+(S \cup \{j\}))$. Moreover, for any party $i$ at round $t$, $\psi_t(i)=0$ if $\forall S \subseteq I_t \setminus \{i\}$, $\psi(I_{1:t-1}+(S \cup \{i\})) = \mathcal{U}(I_{1:t-1}+S)$.  They also extend the previous Shapley value approximation techniques to compute federated Shapley values.

Sim~\textit{et~al.}~\cite{SZCL20} consider the more general situation of collaborative machine learning and advocate using information gain as the utility function. For a model $\theta$ trained on data $D$, the information gain $\mathbb{I}(\theta; D) = \mathbb{H}(\theta)-\mathbb{H}(\theta|D)$, which is the reduction in uncertainty. They generalize to $\rho$-Shapley fairness by assigning a reward $r_i = k \psi_i^\rho$ to a party $i$.  By tuning parameter $\rho$, they can trade off among Shapley fairness, individual rationality, stability of the grand coalition and group welfare.

Hu and Gong~\cite{Hu2020TradingDF} consider privacy leaking in federated learning and design an incentive mechanism to compensate the cost of privacy leakage of the users that are most likely to provide reliable data.  Their problem is formulated in a two-stage Stackelberg game~\cite{Stackelberg}. Richardson~\textit{et~al.}~\cite{richardson2019rewarding} use influence functions to reward data contributions to linear regression in the federated learning setting.

\subsection{Summary}

In this section, we review the topic of pricing data products.  We first analyze the structures, players, and ways to produce data products in data marketplaces.  Then, we examine several important areas in pricing data products, including arbitrage-free pricing, revenue maximization pricing, fair and truthful pricing and privacy preserving pricing. We also discuss how to price dynamic data and online pricing.  When pricing data products in a data marketplace, those several considerations are typically incorporated and integrated in one way or another.

\section{Discussion and Open Challenges}\label{sec:discussion}

\nop{
\subsection{Industry Practice}

DataStreamX information good Pricing Whitepaper, Let's Talk About Data Pricing by Ocean Protocol
}


Data pricing comes from practical demands and has been tackled in multiple disciplines.  Although there is a rich body of literature addressing a series of issues in data pricing, there are still many questions remained unexplored.  In this section, we discuss some interesting challenges for possible future work.  By no means our list is exhaustive.  Instead, we hope our discussion can intrigue more extensive interest and research effort into this fast growing area.

\subsection{Data Supply Chain: A Grand Challenge}

At the macro level, although many studies focus on different steps in data marketplaces, we clearly observe a lack of systematic investigation on data supply chains and development of end-to-end solutions.  As data products are abundant and diversified, to develop ecologically sustainable marketplaces, supply chains of data products have to be built.  Here, we introduce and advocate the notion of \emph{data supply chains}, which connect all parties involved in data production and consumption, including data providers, data processors, data analysts, data product and services consumers and other possible roles. Each party in a data supply chain connects its upstream providers and its downstream consumers, provides its value-added contributions and obtains rewards. Feedback mechanisms through pricing and marketing have to be created in a data supply chain so that supply and consumption can be matched, coordinated and balanced.  Most of those problems are not thoroughly thought about.

Although the notion of data supply chain is not mentioned in literature, some specific trends and challenges are discussed sporadically.  For example, Muschalle~\textit{et~al.}~\cite{10.1007/978-3-642-39872-8_10} identify some trends and challenges in data consumption and marketplaces.  First, they assert that many essential data processing tasks are essential for data markets, such as labeling, annotating and aggregating data.  Second, data markets will be integrated with numerous application domains.  To enable domain data markets, it is important to customize general data processing technologies for niche domains.  Third, customers want to have data faster. Thus, it is important to create online data query services and develop corresponding pricing models.  Fourth, as there are more data, more data providers and more analysts, a data product may be substituted by others. To hatch a healthy ecological data marketplace, it is important to establish standard data processing mashups to facilitate data product substitution.  Fifth, to maintain a fair data market overall, it is important to provide price transparency so that data product providers have to optimize their data and data processing/analysis services. Last, customer preferences and experience are critical for data markets. 

Recently, Acemoglu~\textit{et~al.}~\cite{RePEc:nbr:nberwo:26296} present an insightful study on the ecological effect of data markets.  They demonstrate that a user's sharing of data may likely reveal some other users' privacy and depress the price of other users' data.  The depressed prices lead to excessive data sharing and thus further reduce welfare.  Their study suggests the need of mediation in data sharing in data markets.

Most recently, Fernandez~\textit{et~al.}~\cite{10.14778/3407790.3407800} analyze the challenges and propose a research agenda around constructing a data market platform to address the sharing, discovery and integration of data among many parties.  Their big picture covers both market design and system development.  The focus is to create the incentives and mechanisms to connect data supply and demand.  As the middlemen, arbiters build data mashups to match data supply and demand. The market platforms advocated by the authors can be regarded as the data exchange mechanisms in data supply chain.

One challenge associated with the macro view of data supply chain is the interdisciplinary nature of data pricing research.  As can be observed in this article, data pricing is studied in many different disciplines, such as economics, marketing, electronic commerce, data management, data mining and machine learning.  The communication and dialog among different areas have to be strengthened.

\subsection{Some Technical Challenges at the Micro Level}

At the micro level, there are many research problems remained open. We name a few examples of fundamental problems.  

First, most of the studies suggest relative prices of data products.  Very few studies connect theoretical models with data pricing practice and investigate absolute prices of data products and their marketing effect.  As data pricing is a market mechanism and user behavior in practice is hard to modeled completely, experimental studies of data pricing models are essential and should be connected to theoretical investigations.

Second, pricing is based on valuation and equilibrium among multiple parties. Different parties may have different valuation on data, data products and data services.  It is important to systematically establish the principles of value assessment for various parties in data marketplaces, such as data providers, data owners, data users, and data brokers.  Moreover, it is important to understand what messages are passed to different parties in data marketplaces through data pricing actions, and how. So far, value assessment of data and negotiations among different parties in data marketplaces are largely not analyzed in detail.

Third, many pricing models are proposed in literature. It is important to understand how data pricing models and their assumptions can be implemented and enforced in practice.  Specifically, accounting and auditing in data marketplaces are critical to achieve transparency in data pricing and efficiency in data marketplaces. Accounting and auditing in data marketplaces, however, are interesting problems that have not been investigated in depth yet. We need principles, quality guarantees and designs of operational procedures for accounting and auditing in data pricing, transactions and adversary detection.

Fourth, most of the studies on data pricing develop general models. At the same time, as data science transforms many application domains, data pricing has to deal with specific applications. Mechanisms, regulations and constraints in a specific domain may facilitate data pricing in some aspects, and post challenges in some other aspects.  For example, Jia~\textit{et~al.}~\cite{10.14778/3342263.3342637} show that, although fair pricing in general is exponential in computation time but can be achieved polynomially in kNN models (Section~\ref{sec:fair-pricing}).  It is interesting and highly desirable to explore fairness, truthfulness, and privacy preservation of data pricing in specific applications.

Last but not least, almost all applications are dynamic in nature.  The values of data, data products and data services may also evolve over time.  The changes may be caused by the updates in demands and supplies.  It is important to develop mechanisms to capture and monitor changes in demand and supply of data, data products and data services, and explore corresponding dynamic pricing.



\bibliographystyle{IEEEtranS}
\bibliography{ref-url}

\nop{
\begin{IEEEbiography}\begin{IEEEbiography}[{\includegraphics[width=1in,height=1.25in,clip,keepaspectratio]{./pei.jpg}}]{Jian Pei}
(Fellow) is a Professor at the School of
Computing Science at Simon Fraser University,
Canada. His research interests can be
summarized as developing effective and effi-
cient data analysis techniques for novel data
intensive applications. He is currently interested
in various techniques of data mining,
Web search, information retrieval, data warehousing,
online analytical processing, and
database systems, as well as their applications
in social networks, health-informatics,
business and bioinformatics. His research has been supported in
part by government funding agencies and industry partners. He has
published prolifically and served regularly for the leading academic
journals and conferences in his fields. He is an associate editor of
ACM Transactions on Knowledge Discovery from Data (TKDD). He is a Fellow of of the Association
for Computing Machinery (ACM) and the Institute of Electrical and
Electronics Engineers (IEEE). He is the recipient of several prestigious
awards.
\end{IEEEbiography}
}

\end{sloppy}
\end{document}